\documentclass{jfm}

\usepackage{graphicx}
\usepackage{natbib}

\usepackage{amssymb}
\usepackage{amsmath}
\usepackage{subfig}

\usepackage{color} % for annotations

\newcommand{\mr}[1]{\mathrm{#1}}
\newcommand{\pd}{\partial}
\newcommand{\lk}{\left}%linke Klammer
\newcommand{\rk}{\right}%rechte Klammer
\newcommand{\dd}{\mathrm{d}}
\newcommand{\ii}{\mathrm{i}}

\newcommand{\fz}{\mathfrak{z}}
\newcommand{\lnppp}{\ln\bigg(\frac{1+\sin\left(\frac{\pi}{2}a\right)}{1-\sin\left(\frac{\pi}{2}a\right)}\bigg)}
\newcommand{\erf}{\mathrm{erf}}

\newcommand{\und}{\quad \text{and}\quad}
\newcommand{\for}{\quad\text{for}\quad}
\newcommand{\reel}{\mathrm{Re}}
\newcommand{\imag}{\mathrm{Im}}

\ifCUPmtlplainloaded \else
  \checkfont{eurm10}
  \iffontfound
    \IfFileExists{upmath.sty}
      {\typeout{^^JFound AMS Euler Roman fonts on the system,
                   using the 'upmath' package.^^J}%
       \usepackage{upmath}}
      {\typeout{^^JFound AMS Euler Roman fonts on the system, but you
                   dont seem to have the}%
       \typeout{'upmath' package installed. JFM.cls can take advantage
                 of these fonts,^^Jif you use 'upmath' package.^^J}%
      }
  \else
  \fi
\fi

\ifCUPmtlplainloaded \else
  \checkfont{msam10}
  \iffontfound
    \IfFileExists{amssymb.sty}
      {\typeout{^^JFound AMS Symbol fonts on the system, using the
                'amssymb' package.^^J}%
       \usepackage{amssymb}%
         \let\leq=\leqslant
         
      }{}
  \fi
\fi

\ifCUPmtlplainloaded \else
  \IfFileExists{amsbsy.sty}
    {\typeout{^^JFound the 'amsbsy' package on the system, using it.^^J}%
     \usepackage{amsbsy}}
    {}
\fi

\newcommand\Rey{\mbox{\textit{Re}}}  % Reynolds number
\newsavebox{\astrutbox}
\sbox{\astrutbox}{\rule[-5pt]{0pt}{20pt}}

\title[Flow over a microstructured surface in the Cassie state]{Influence of the enclosed fluid on the flow over a microstructured surface in the Cassie state}

\author[C. Sch\"onecker , T. Baier
and S. Hardt]%
{C\ls L\ls A\ls R\ls I\ls S\ls S\ls A\ns S\ls C\ls H\ls \"O\ls N\ls E\ls C\ls K\ls E\ls R%$^{1, \, 2}$%
  \thanks{Email address for correspondence: schoenecker@csi.tu-darmstadt.de
  },\ns
T\ls O\ls B\ls I\ls A\ls S\ns B\ls A\ls I\ls E\ls R\break
\and S\ls T\ls E\ls F\ls F\ls E\ls N\ns H\ls A\ls R\ls D\ls T}

\affiliation{Institute for Nano- and Microfluidics, Center of Smart Interfaces, Technische Universit\"at Darmstadt, 64287 Darmstadt, Germany\\[\affilskip]
}

\pubyear{2010}
\volume{650}
\pagerange{119--126}
\date{?; revised ?; accepted ?. - To be entered by editorial office}
\begin{document}

\maketitle

\begin{abstract}
Analytical expressions for the flow field as well as for the effective slip length of a shear flow over a surface with periodic rectangular grooves are derived. The primary fluid is in the Cassie state with the grooves being filled with a secondary immiscible fluid. The coupling of both fluids is reflected in a locally varying slip distribution along the fluid-fluid interface, which models the effect of the secondary fluid on the outer flow. The obtained closed-form analytical expressions for the flow field and effective slip length of the primary fluid explicitly contain the influence of the viscosities of the two fluids as well as the magnitude of the local slip, which is a function of the surface geometry. They agree well with results from numerical computations of the full geometry. The analytical expressions allow investigating the influence of the viscous stresses inside the secondary fluid for arbitrary geometries of the rectangular grooves. For classic superhydrophobic surfaces, the deviations in the effective slip length compared to the case of inviscid gas flow are are pointed out. Another important finding with respect to an accurate modeling of flow over microstructured surfaces is that the local slip length of a grooved surface is anisotropic. %So far, it has been modeled as isotropic. 
\end{abstract}


\section{Introduction} 

Flow over micro- or nanostructured surfaces, whose indentations are filled by a second immiscible fluid can be encountered in a number of applications. In the case of classic superhydrophobic surfaces, air is enclosed in between the water flowing over the surface and the surface itself. This situation is the key to many typical properties of such surfaces, like the inhibited wetting or the easy slipping of water along the surface. Other examples of flow along structured surfaces include porous media or membranes, which may be filled with an arbitrary secondary fluid. Generally, all situations where a fluid is in the Cassie state can be considered. In all these cases, the characteristics of the microstructures and the properties of the secondary fluid have a specific effect on the primary flow. 

Usually, the surface indentations are numerous and not of the same scale as the overall geometry considered. These circumstances generally make a full numerical solution intractable and call for an effective boundary condition that describes the net effect of the structured surface on the flow.
This boundary condition appears in the form of a Navier slip condition
\begin{equation}
\left.u\right|_\mr{boundary}=\beta_\mr{eff}\left.\frac{\pd u}{\pd y}\right|_\mr{boundary},
\label{eq:navierbceff}
\end{equation}
where the effective slip length $\beta_\mr{eff}$ represents the overall effect of the underlying surface on flow above it. In a way, the effective slip length can be understood to be inherent to the surface, including the fluid properties. Therefore, if determined once, it can be used in analytical calculations or numerical simulations, dramatically reducing their complexity and computational costs or making certain problems accessible at all. All these points make the effective slip length a parameter of utmost interest. However, explicit expressions for $\beta_\mr{eff}$ are known only for a very limited number of cases. 

The oldest and best known expressions are due to \citeauthor{philip1972} (\citeyear{philip1972, philip1972b}). He considered flow over mixed no-slip and no-shear conditions for various surface geometries. In the case of flow over a periodic array of no-shear stripes within a flat surface, he obtained for the transverse and longitudinal effective slip length
\begin{align}
\beta_\mr{t,\,P}&=-\frac{L}{2 \pi}\ln\lk(\cos\lk(\frac{\pi}{2}a\rk)\rk),\\
\beta_\mr{l,\,P}&=-\frac{L}{\pi}\ln\lk(\cos\lk(\frac{\pi}{2}a\rk)\rk),
\label{eq:betaphilip}
\end{align}
with $L$ being the period of the no-shear stripes and $a$ the corresponding free-interface fraction. In this case, the longitudinal effective slip length is simply twice the transverse one. The no-shear condition is equivalent to a fluid of vanishing viscosity filling the surface corrugations. This provides an infinite local slip at the fluid-fluid interface \citep[cf.][]{eijkel2007}. Although such a fluid is obviously nonexistent, this assumption has been widely used in the modeling of water-air interfaces, for instance at superhydrophobic surfaces \citep{lauga2003, sbragaglia2007, Davis2009, crowdy2010, davies2010}. Indeed, the viscosity of air is very low compared to most other fluids, which makes the assumption reasonable in many practical cases. Even so, it has never been really verified. A no-shear boundary also implies that the structure of the surface underneath the fluid-fluid interface is insignificant. This might be true as long as this structure is large enough. For very thin films of the lubricating fluid, the thickness of the film may have an influence on the velocity profile close to the fluid-fluid interface. With the ongoing miniaturization of technological systems, the assumption of no-shear fluid-fluid interfaces becomes questionable.

The problem of a finite local slip at the fluid-fluid interface has been addressed by \citet{belyaev2010}. They consider a superhydrophobic surface as an array of stripes of constant local slip within a no-slip wall. A Fourier expansion of these boundary conditions leads to a so-called problem of dual trigonometric series \citep{sneddon_b_mixedboundary}. \citeauthor{belyaev2010} calculate the corresponding transverse and longitudinal effective slip length to be
\begin{align}
\beta_\mr{t,\,BV}&=-\frac{L}{2 \pi}\frac{\ln\lk(\cos(\frac{\pi}{2}a)\rk)}{1+\frac{L}{2\pi b_\mr{BV}}\ln\lk(\sec(\frac{\pi}{2}a)+\tan(\frac{\pi}{2}a)\rk)}, \label{eq:betabvt} \\ 
\beta_\mr{l,\,BV}&=-\frac{L}{\pi}\frac{\ln\lk(\cos(\frac{\pi}{2}a)\rk)}{1+\frac{L}{\pi b_\mr{BV}}\ln\lk(\sec(\frac{\pi}{2}a)+\tan(\frac{\pi}{2}a)\rk)}.\label{eq:betabvl}
\end{align}
Here, the constant local slip is called $b_\mr{BV}$, and $a$ correspondingly is the fraction of the slipping interface. Clearly, Philip's results are an integral part of \citeauthor{belyaev2010}'s equations. The corresponding relation between the transverse and longitudinal effective slip lengths, which has been investigated by \citet{asmolov2012}, is not a simple doubling any more. 
As can be seen from the above equations, the transverse effective slip length can be obtained from the longitudinal equation by doubling the local slip and dividing the whole expression by a factor of two. 

Equations \eqref{eq:betabvt} and \eqref{eq:betabvl} are supposed to apply to superhydrophobic surface, however, it is not yet clear what value to choose for $b_\mr{BV}$ for specific surfaces. Besides this, the local slip length at a superhydrophobic surface is rather expected to be a continuous function than a step-like function, exhibiting a jump in the local slip at the edges of the stripes. The slip of a superhydrophobic surface is actually an apparent slip (in the terminology of \citet{lauga_handbook}), which is determined by the hydrodynamics of the enclosed gas. The local behavior is therefore continuous, which has already been shown in the case of flow over a single groove filled with an arbitrary secondary fluid \citep{schoenecker2013}. A jump in the local slip length at the edges of the grooves could only be provoked by an intrinsic slip, which is only of minor influence here.

Besides conventional superhydrophobic surfaces, other functional microstructured surfaces are currently under development. Instead of air being enclosed in the corrugations, also other materials like for instance oil could be employed, which promises an improved stability of the Cassie state against pressure \citep{wong2011} or enhanced electrical properties \citep{steffes2011}. The choice of an appropriate groove-filling medium provides the chance to specifically tailor the surface properties, for examples in terms of phobicity and philicity, not only with regard to water, but to any desired fluid. 

In all these cases, the viscosities of the involved fluids are even more important than for a conventional superhydrophobic surface. Different viscosities should lead to a different local slip. In previous work, this effect has at most be treated qualitatively based on scaling arguments \citep{vinogradova1995, degennes2002, ybert2007}. It has not yet been described how viscosities contribute to the effective slip length of a structured surface.

If viscosity influences the effective slip length, so does the geometry of the surface structure. As already mentioned above, even for a gas filling the corrugations the geometry should affect the effective slip length anyway in certain cases, like those leading to a thin film of the lubricating fluid. The geometry influence is a further subject, which has not yet been systematically investigated. 

Apart from the above mentioned work, other investigations of the effective slip length are mostly based on Fourier expansions, which have only been solved numerically \citep{wang2003, ng2010, ng2011}. They are generally restricted to perfect or constant slip and only provide asymptotic considerations or fits for the effective slip length in certain limiting cases as small free surface fractions (e.g. \citet{sbragaglia2007b}) or small solid interface fractions (e.g. \citet{ng2010}). Hence a general theory, providing an explicit expression for the effective slip length as well as for the flow field and taking into account the influence of viscosity and surface structure is still lacking. 

\begin{figure}
  \centerline{\includegraphics[width=0.6\textwidth]{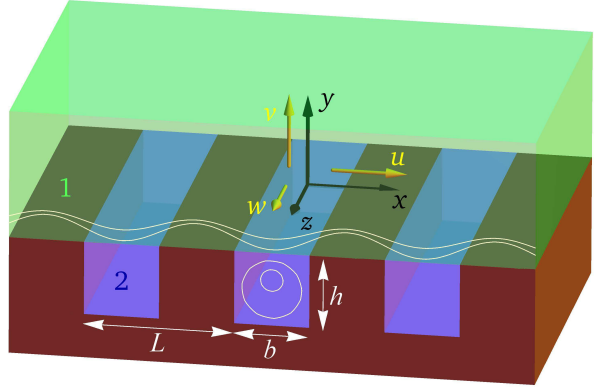}}
  \caption{Schematic of the microstructured surface with coordinate and velocity directions. Fluid 1 is above the surface, while fluid 2 fills the grooves. Streamlines exemplarily illustrate flow transverse to the grooves.}
\label{fig:shp_oberfl}
\end{figure}

In this work, the flow over a periodic array of rectangular grooves as depicted in fig. \ref{fig:shp_oberfl} shall be considered. The grooves of width $b$, depth $h$ and period $L$ are filled with a fluid being immiscible with the fluid flowing over them. Analytical expressions for both the flow field and the effective slip length are derived for transverse and longitudinal flow with respect to the grooves. The effect of the grooves on the outer flow is modeled as a local slip, taking into account the viscosity ratio of the two fluids as well as the aspect ratio of the grooves. The results are relatively simple, yet accurate, closed-form equations, which predict the flow field and the effective slip length for any fluid-fluid interface fraction, viscosity ratio of the fluids and aspect ratio of the grooves. Moreover, these equations directly lead to several further findings, like a mathematical relation between transverse and longitudinal flow.

%%%%%%%%%%%%%%%%%%%%%%%%%%%%%%%%%%%%%%%%%%%%%%%%%%%%%%%%%%%%%%%%%%%
%%%%%%%%%%%%%%%%%%%%%%%%%%%%%%%%%%%%%%%%%%%%%%%%%%%%%%%%%%%%
\section[Mathematical description of the flow field and the effective slip length]{Mathematical description of the flow field and the effective slip length}\label{matheperiodisch}
In the following, first the basic mathematical structure describing the flow of fluid 1 over the surface is introduced. The results are exact solutions to a specific local slip-length distribution, which models the flow over a periodic array of rectangular grooves. The actual magnitude of the local slip length will be modeled in a second step. Both parts together yield a full description of the shear flow of a fluid moving over a periodic array of rectangular grooves filled with a second immiscible fluid. The results provide insight into the effective slip length of such surfaces in particular as well as into the calculation of the effective slip of patterned surfaces in general.

%%%%%%%%%%%%%%%%%%%%%%%%%%%%%%%%%
\subsection{Transverse flow}
As the first fundamental case, flow in transverse direction over an array of rectangular cavities of period $L$ is considered. It is assumed that the Reynolds number is small enough to describe the flow by the Stokes equations. Then, the streamfunction $\Psi$, being related to the velocities $u$ and $v$ in the $x$ and $y$ directions, respectively, via
\begin{equation}
\frac{\pd \Psi}{\pd x}=-v \und \frac{\pd \Psi}{\pd y}=u,
\label{eq:stromfkt}
\end{equation}
obeys the biharmonic equation
\begin{equation}
\Delta \Delta \Psi=0.
\label{eq:biharmonisch2}
\end{equation}
The flow shall be driven by a shear stress $\tau_\infty$ in the $x$ direction at $y\rightarrow\infty$, where the variations in the flow owing to the spatially changing boundary conditions will have equilibrated. 
Due to the symmetry of the problem, it is sufficient to look for a solution in the interval $0\leq x\leq L/2$, where $L$ is the period of the grooves. The boundary conditions at the surface read
\begin{align}
u&=N \gamma_\mr{t}(x)\frac{\pd u}{\pd y} &\for 0\leq x &\leq\frac{b}{2},\quad y=0 \label{eq:bcreel1p}\\
u&=0 &\for \frac{b}{2}< x &\leq\frac{L}{2},\quad y=0  \label{eq:bcreel2p}
\end{align}
with the viscosity ratio of the fluids 1 and 2
\begin{equation}
N=\frac{\eta_1}{\eta_2}
\label{eq:N}
\end{equation}
and the local slip-length distribution $\gamma_\mr t (x)$, representing the effect of the fluid movement in the grooves. Equation \eqref{eq:bcreel1p} can be understood as a Navier slip condition with a spatially varying local slip-length distribution 
\begin{equation}
\Gamma_\mr{net, local}(x)=N\gamma_\mr t(x).
\label{eq:nettoslip}
\end{equation}
The boundary condition follows from the continuity of velocity and shear stress between both fluids at the fluid-fluid interface. The flow velocity and the shear stress of fluid 2 at the interface fulfill a relationship that can be interpreted as a Navier slip condition 
\begin{equation}
u_2=\gamma_\mr t(x) \frac{\pd u_2}{\pd y}.
\label{eq:slip2}
\end{equation}
The specific form of the local slip $\gamma_\mr t (x)$ will follow from a consideration of its origin, that is the flow inside the groove. At this point, we may want to distinguish between molecular and apparent slip (for a review, e.g. see \citet{lauga_handbook}). The first one describes slip on the molecular level and is of the order of nanometers, while the latter may be due to effects like an enclosed gas, resulting in a local slip length of the order of the length scale of the surface features. Here it is assumed that the surface features are much larger than any hypothetical molecular slip length. Due to this difference in magnitude, a situation is considered, where the molecular slip is negligible. The apparent slip however is entirely described by the hydrodynamics of the fluid inside the grooves, and therefore accessible to an analysis. The concept of apparent slip is fully consistent with the continuity of the velocity field across the fluid-fluid interface, i.e. it is only an auxiliary concept, which is employed to compute the flow above the grooves. Its specific description is similar to previous work on the flow over a single groove \citep{schoenecker2013}. In this case, an elliptical function of amplitude $d_\mr t$ described the local slip-length distribution $\gamma_\mr t (x)$ reasonably well, where $d_\mr t$ was merely a function of the groove aspect ratio. In the current periodic case, an extended description has to be found. The no-slip condition at the wall requires $\gamma_\mr t(b/2)=0$. For Stokes flow, the local slip length distribution furthermore takes a maximum value $d_\mr t$ at the center of the cavities, i.e. $\gamma_\mr t (0)=d_\mr t$. So far, micro- or nanostructured surfaces have mainly been modeled on a rather abstract level, employing an either finite or infinite constant local slip. By considering the actual origin of the slip, the present consideration introduces a modeling of the local slip that represents its distribution in realistic situations much more closely.

To this aim, a general solution for the flow field and the effective slip length due to a specific local slip-length distribution $\gamma_\mr t (x)$ is derived. Later, the magnitude $d_\mr t$ of this distribution 
 will be considered.

%%%%%%%%%%%%%%%%%%%%%%%%%%%%%%%
\subsubsection{Flow field}

The model equations \eqref{eq:stromfkt}--\eqref{eq:bcreel2p} can be reformulated using complex analysis. A general solution to the biharmonic equation \eqref{eq:biharmonisch2} is given by the Goursat theorem. For a flow over an impermeable wall, it takes the form \citep[see][]{muskhelishvili_b_elasticity,garabedian1966,philip1972}
\begin{equation}
\Psi=\mathrm{Re}\left((\bar{\mathfrak{z}}-\mathfrak{z})W_\mr{t}(\mathfrak{z})\right)
\label{eq:Goursat}
\end{equation}
with $\mathrm{Re}$ indicating the real part. The complex coordinate is
\begin{equation}
\fz=x+\ii y \und \bar \fz=x-\ii y.
\label{eq:z}
\end{equation}
The Goursat function $W_\mr t(\fz)$ is an analytic function to be determined.

In the boundary condition \eqref{eq:bcreel1p}, the local slip-length distribution $\gamma_\mr t(x)$ is to be specified. 
It has been shown that for a shear flow over a single groove the fluid-fluid interface can be modeled by a constant-shear-stress boundary condition \citep{schoenecker2013}. 
For the single groove, this condition is equivalent to an elliptic local slip-length distribution. Correspondingly, flow over a periodic array of grooves is modeled as an array of constant-shear slots. In this case, the flow field can be described by a simple superposition technique. The boundary conditions of constant shear stress at the fluid-fluid interface, no slip at the wall and a given shear stress far from the surface can be fulfilled with a superposition of a plain Couette flow and \citeauthor{philip1972}'s (\citeyear{philip1972}) solution for mixed no-shear and no-slip conditions. 

In terms of the Goursat function $W_\mr{t}(\fz)$ the superposition ansatz reads
\begin{equation}
W_\mr{t}(\fz)=C_1 \fz+C_2 \arccos\left(\frac{\cos\big(\frac{\pi \fz}{L}\big)}{\cos\big(\frac{\pi}{2}a\big)}\right),
\label{eq:superposansatzt}
\end{equation}
with two constants $C_1$ and $C_2$ to be determined. The fluid-fluid interface fraction is denoted by $a$. The first term of eq. \eqref{eq:superposansatzt} represents a plain shear flow and contributes a constant shear stress everywhere on $y=0$, while the second term is essentially \citeauthor{philip1972}'s (\citeyear{philip1972}) expression for a periodic array of no-shear slots between no-slip stripes. Under the conditions that the combined flow is driven by a shear stress $\tau_\infty$ far away from the surface and that the local slip length takes the value $d_\mr t=\frac{1}{N}\frac{u}{\pd u/\pd y}$ at the center of the cavities, both constants are readily determined.

The resulting solution for the transverse stream function is
\begin{equation}
\Psi=\frac{\tau_\infty}{2 \eta_1}\Bigg(y^2+\frac{-\pi y^2+ y L\,\mr{Im}\left(\mr{arccos}\bigg(\frac{\cos\lk(\frac{\pi\fz}{L}\rk)}{\cos\lk(\frac{\pi}{2}a\rk)}\bigg)\right)}{\pi+\frac{1}{4 a D_\mr{t} N}\lnppp%\left(-1-\frac{2}{-1+\sin\lk(\frac{\pi}{2}a\rk)}\right)
}\Bigg)
\label{eq:stromfkttransperiod}
\end{equation}
where $D_\mr t=d_\mr t/b$ is the nondimensionalized maximum local slip length. With the normalized variables $\tilde \Psi=\Psi \frac{8\eta_1}{\tau_\infty b^2}$, $X=x \frac{b}{2}$ and $Y=y \frac{b}{2}$, the nondimensionalized form reads
\begin{equation}
\tilde\Psi=\frac{Y^2+C_\mr t \,Y \alpha^{-1} \,\mr{Im}\left(\mr{arccos}\lk(\frac{\cos\lk(\alpha(X+\mr i Y)\rk)}{\cos(\alpha)}\rk)\right)}{1+C_\mr t}
\label{eq:stromfkttransperioded}
\end{equation}
abbreviating $\alpha=\frac{\pi}{2}a$ (analogously to \citet{philip1972}) and 
\begin{equation}
C_\mr t=\frac{4\pi a D_\mr{t} N}{\lnppp}.%\left(-1-\frac{2}{-1+\sin\lk(\frac{\pi}{2}a\rk)}\right)}.
\label{eq:alphaundc}
\end{equation}

With equation \eqref{eq:stromfkttransperioded}, a closed-form analytical expression for the flow over a periodic array of patches with a local slip length $\gamma_\mr t(x)$ is available. 
The constant-shear modeling of the cavities relates to a local slip-length distribution of
\begin{equation}
	\begin{split}
	\frac{\gamma_\mr t(x)}{b}&=\frac{1}{b}\frac{1}{N}\frac{u(x)}{\pd u(x)/\pd y}\\
	&=\frac{1}{4 N} C_\mr t \alpha^{-1} \mr{Im}\bigg(\mr{arccos}\bigg(\frac{\cos(\alpha X)}{\cos(\alpha)}\bigg)\bigg)\\
	&=2 D_\mr t \frac{\mr{Im}\bigg(\mr{arccos}\bigg(\frac{\cos\lk(\frac{\pi x}{L}\rk)}{\cos\lk(\frac{\pi}{2}a\rk)}\bigg)\bigg)}{\lnppp},
	\end{split}
	\label{eq:gammat}
\end{equation}
ensuring also $\gamma_\mr t(b/2)=0$. The distribution is characterized by a maximum value $D_\mr t$. 
This form of the local slip-length distribution represents a generalization of the single-groove case, involving the fluidic surface fraction $a$. As $a\rightarrow 0$, i.e. in the single-groove limit, it exactly takes the elliptic form reported in \citet{schoenecker2013}. %The slip-length distribution is suitable for modeling the behavior of a periodic array of rectangular cavities as will be illustrated later. 

It should be noted that the constant shear stress model and the resulting slip-length distribution represent assumptions that will be tested in subsequent sections. In a more rigorous sense this model can be viewed as a first step in an iteration scheme allowing to determine successively improved boundary conditions representing the coupling between fluid 1 and fluid 2, as outlined by \citet{schoenecker2013}.

%%%%%%%%%%%%%%%%%%%%%%%%%%%%%%%%%%%%%%
\subsubsection{Effective slip length}

The analytical expressions for the flow field provide all information to determine the effective slip length. 
For this purpose we consider a uniform flow, which behaves like a plain shear flow along a wall with some effective slip $\beta_t$
\begin{equation}
\bar u(y) =\frac{\tau_\infty}{\eta_1}(y+\beta_\mr t).
\label{eq:shearslipt}
\end{equation}
Requiring this shear flow to obtain the same velocity far from the wall as a flow over a structured surface $u(x,y)$, linearity of the underlying equations leads to the condition
\begin{equation}
\beta_\mr t=\frac{\eta_1}{\tau_\infty}\bar u(0),
\label{eq:sliptgemittelt}
\end{equation}
where
\begin{equation}
\bar u(0)=\frac{1}{L}\int\limits_{-L/2}^{L/2} u(x,0)\, \dd x
\label{eq:umittel}
\end{equation}
is the mean velocity at $y=0$ \citep[for a derivation, see][]{squires2008}.

To evaluate the integral in eq. \eqref{eq:umittel} it is necessary to use the integral relations derived by \citet{philip1972b}, however, with the superposition technique employed here, the effective slip length can be determined in an even more straightforward manner. Equation \eqref{eq:stromfkttransperioded} shows that the shear flow part does not contribute to the slip length, since its velocity at $y=0$ is zero, while the Philip part contributes with a factor of $C_\mr t/(1+C_\mr t)$. The normalized slip length $\beta^*_\mr t=\beta_\mr{t}/ L$ can therefore directly be written as 
\begin{equation} \label{eq:betatedc}
	\begin{split}
\beta^*_\mr t &=\frac{C_\mr t}{1+C_\mr t}\,\beta_\mr{t,\, P}^*\\
&=-\frac{1}{2\pi}\frac{C_\mr t}{1+C_\mr t}\ln(\cos(\alpha)),
	\end{split}
\end{equation}
which is
\begin{equation}
	\beta^*_\mr t=-\frac{\ln\big(\cos(\frac{\pi}{2}a)\big)}{2 \pi +\frac{1}{2 a D_\mr t N}\,\lnppp}.%(-1-\frac{2}{-1+\sin(\frac{\pi b}{2 L})})}.
	\label{eq:betated}
\end{equation}

%%%%%%%%%%%%%%%%%%%%%%%%%%%%%%%%%%%%%%%%%%%%%%%%%%%%%%%%%%%%%%%%%
\subsection{Longitudinal flow}

In the case of fluid flowing longitudinally over the grooves, the governing equation for the flow field is the Laplace equation
\begin{equation}
 \frac{\pd^2 w}{\pd x^2}+\frac{\pd^2 w}{\pd y^2}=0
\label{eq:laplacep}
\end{equation}
with the boundary conditions
\begin{align}
w&=N \gamma_\mr{l}(x)\frac{\pd w}{\pd y} &\for 0\leq x &\leq\frac{b}{2},\quad y=0 \label{eq:bcreel1lp}\\
w&=0 &\for \frac{b}{2}< x &\leq\frac{L}{2},\quad y=0.  \label{eq:bcreel2lp}
\end{align}
Correspondingly, the flow is driven by a shear stress $\tau_\infty$ in the $z$ direction at $y\rightarrow\infty$. Due to periodicity only the interval between $x=0$ and $x=L/2$ is considered. 

An analytic function $W_\mr l(\mathfrak{z})$ with the choice
\begin{equation}
w=\imag(W_\mr l(\mathfrak{z}))
\label{eq:wWp}
\end{equation}
fulfills the Laplace equation. Following the same superposition strategy as employed in the transverse case, a solution to the above problem can be written as
\begin{equation}
W_\mr{l}(\fz)=C_3 \fz+C_4 \arccos\left(\frac{\cos\big(\frac{\pi \fz}{L}\big)}{\cos\big(\frac{\pi b}{2 L}\big)}\right).
\label{eq:superposansatzl}
\end{equation}
Determining the constants $C_3$ and $C_4$ such that the shear stress is $\tau_\infty$ at $y\rightarrow\infty$ and that the normalized slip length is $D_\mr l=d_\mr l/b$ at $\fz=0$, the velocity becomes
\begin{equation}
w=\frac{\tau_\infty}{\eta_1}\Bigg(y+\frac{-\pi y+ y L\,\mr{Im}\left(\mr{arccos}\bigg(\frac{\cos\lk(\frac{\pi\fz}{L}\rk)}{\cos\lk(\frac{\pi}{2}a\rk)}\bigg)\right)}{\pi+\frac{1}{2 a D_\mr{l} N}\lnppp%\left(-1-\frac{2}{-1+\sin(\frac{\pi}{2}a)}\right)
}\Bigg),
\label{eq:wlongp}
\end{equation}
or in a notation, where the mathematical principles and the correspondence to the transverse case are more obvious
\begin{equation}
\tilde w =\frac{Y+C_\mr l \,\alpha^{-1} \,\mr{Im}\left(\mr{arccos}\lk(\frac{\cos\lk(\alpha(X+\mr i Y)\rk)}{\cos(\alpha)}\rk)\right)}{1+C_\mr l}
\label{eq:wlongperioded}
\end{equation}
with $\tilde w=w \frac{2 \eta_1}{\tau_\infty b}$ and
\begin{equation}
C_\mr l=\frac{2\pi a D_\mr{l} N}{\lnppp%\left(-1-\frac{2}{-1+\sin(\frac{\pi}{2}a)}\right)
}.
\label{eq:Clong}
\end{equation}
The local slip-length distribution 
\begin{equation}
	\begin{split}
	\frac{\gamma_\mr l(x)}{b}	&=\frac{1}{2 N} C_\mr l \alpha^{-1} \mr{Im}\bigg(\mr{arccos}\bigg(\frac{\cos\lk(\alpha X\rk)}{\cos\lk(\alpha\rk)}\bigg)\bigg)\\
	&=2 D_\mr l \frac{\mr{Im}\bigg(\mr{arccos}\bigg(\frac{\cos(\frac{\pi x}{L})}{\cos(\frac{\pi}{2}a)}\bigg)\bigg)}{\lnppp}
	\end{split}
	\label{eq:gammal}
\end{equation}
is also analogous to the transverse case.

Eventually, the effective longitudinal slip length is found as
\begin{equation} \label{eq:betaledc}
	\begin{split}
\beta^*_\mr l &=\frac{C_\mr l}{1+C_\mr l}\,\beta_\mr{l,\,P}^*\\
&=-\frac{\ln\big(\cos(\frac{\pi}{2}a)\big)}{\pi +\frac{1}{2 a D_\mr l N}\,\lnppp%(-1-\frac{2}{-1+\sin(\frac{\pi b}{2 L})})
}.
	\end{split}
\end{equation}

%%%%%%%%%%%%%%%%%%%%%%%%%%%%%%%%%%%%%%%%%%%%%%%%%%%%%%%%%%%%%%%%%
%%%%%%%%%%%%%%%%%%%%%%%%%%%%%%%%%%%%%%%%%%%%%%%%%%%%%%%%%%%%%%%%%
\section{Modeling of the maximum local slip length} \label{slipmodel}

It now remains to relate the local slip length to the specific surface structure. While the equations given so far are the exact response of an outer flow to a specific slip-length distribution, it is crucial to furthermore investigate the dependence of this slip-length distribution on the underlying geometry, that is to model $D_\mr t$ and $D_\mr l$ as a function of the groove aspect ratio of a rectangular groove and the fluidic interface fraction. At low Reynolds numbers of the flow inside the groove, the flow pattern of fluid 2 can be considered as independent of the viscosity ratio. It only depends on the surface geometry. For a single groove, the groove aspect ratio $A$ is the main parameter of interest. For a periodic array of grooves, $D_\mr t$ and $D_\mr l$ are expected to be functions of the aspect ratio $A$ as well as the fluidic interface fraction $a$. If $a$ approaches 1, the flow can no longer be regarded as resulting from a superposition of independent cavities, but the cavities may have an effect on each other. At the same time, this regime is of great practical interest, as a high fluidic interface fraction naturally provides the highest effective slip. Consequently, it is important to adequately address this aspect. In the following, the maximum local slip length shall be parameterized as a function of $A$ and $a$, starting with the longitudinal flow direction. Here, more appropriate analytical expressions are known than for the transverse direction, although the physical foundations of the parameterization are somewhat more comprehensible in the latter case.

\subsection{Longitudinal flow}

Based on investigations of lid-driven cavity flows \citep{pan1967}, it has been shown that the dependence of the maximum local slip length $D_\mr l$ on the aspect ratio $A$ is governed by two physical regimes \citep{schoenecker2013}. At low aspect ratios, $D_\mr l$ increases nearly linearly with $A$, while reaching a plateau value when the cavity or groove becomes deeper than wide. For flow over a single groove, this behavior has been described well by an error function. Accordingly, the corresponding ansatz for the periodic case is
\begin{equation}
D_\mr l(A,a)=d_{0,\mr l}(a)\, \erf\lk(d_{1,\mr l}(a)\,A\rk), 
\label{eq:ansatzdlp}
\end{equation}
where the two coefficients $d_{0,\mr l}(a)$ and $d_{1,\mr l}(a)$ may now depend on the fluidic interface fraction. The former determines the plateau value at $A\rightarrow\infty$, while, the latter controls the slope at low $A$.

\subsubsection{Plateau regime}

The height of the plateau extending from around $A=1$ to $A\rightarrow\infty$ can be determined from known expressions for the effective slip length in this limit. For any known $\beta_\mr l^*$, equation \eqref{eq:betaledc} then yields the corresponding condition for $D_\mr l$.

When investigating flow over porous materials, \citet{richardson1971} reported the effective slip length for longitudinal single-phase flow over a periodic array of infinitely deep grooves as
\begin{equation}
\lk.\beta_\mr {l,\,1}^*\rk|_{A \rightarrow \infty}=\frac{1}{2 \pi}\lk((1+a)\ln(1+a)+(1-a)\ln(1-a)\rk).
\label{eq:richardson}
\end{equation}
For longitudinal flow, it has been assumed that the flow is symmetric in the $z$ direction, meaning that the $w$ velocity is decoupled from $u$ and $v$. The longitudinal effective slip length values for a two-phase flow with $N=1$ and single-phase flow are therefore equivalent and the above expression is fully transferable to the present situation. 

Equations \eqref{eq:richardson} and \eqref{eq:betaledc} yield a functional dependence of the maximum longitudinal slip length on the fluidic interface fraction at $A\rightarrow\infty$, stating
\begin{equation}
\left.D_\mr l\right|_{A\rightarrow\infty}=-\frac{\lnppp}{2 \pi a\bigg(1+\frac{\ln\lk(\cos(\frac{\pi}{2}a)\rk)}{a \mr{arctanh}(a)+\ln\lk(1-a^2\rk)}\bigg)}=f(a)\frac{\ln 2}{\pi}.
\label{eq:Dla}
\end{equation}
The magnitude of the maximum effective slip length for $N=1$ is 
\begin{equation}
\beta_{\mr{ l,\,N=1,\,max}}^*=\lk.D_\mr l\right|_{a=1,\,A\rightarrow\infty}=\frac{\ln 2}{\pi},
\label{eq:dlmax}
\end{equation}
and the variation of $D_\mr l$ relative to its value at $A\rightarrow\infty$ is abbreviated as $f(a)$. Substitution of \eqref{eq:Dla} into the ansatz \eqref{eq:ansatzdlp} directly leads to the first coefficient 
\begin{equation}
d_{0,\mr l}(a)=f(a)\, \beta_{\mr{ l,\,N=1,\,max}}^*.
\label{eq:d0l}
\end{equation}

\subsubsection{Low-aspect-ratio regime}

The second coefficient $d_{1,\mr l}(a)$ can be related to the behavior at low aspect ratios, specifically to the slope of $D_\mr l$ with respect to $A$ at $A\rightarrow 0$.

\citet{bechert1989} discovered that the dependence of the longitudinal effective slip length of a single-phase flow over an array of infinitely thin plates on the depth of the plates is given by
\begin{equation}
\lk.\beta_\mr l^*\rk|_{a=1,N=1}=\frac{1}{\pi}\ln\lk(1+\tanh(\pi A)\rk).
\label{eq:bechert}
\end{equation}

This expression coincides with eq. \eqref{eq:dlmax} if $A\rightarrow\infty$. 
Hence, with eq. \eqref{eq:betaledc} it immediately follows that at $a=1$
\begin{equation}
\lim\limits_{A\rightarrow 0}\lk.\frac{\dd D_\mr l}{\dd A}\rk|_{a=1}=1.
\label{eq:dabed1}
\end{equation}
In the opposing limit of $a\rightarrow 0$, i.e. a single groove, the very shallow groove ($A\rightarrow 0$) can be compared with a film flow in the lubrication limit. The lubrication approximation gives a slope of 1 for the local slip length as a function of $A$ at $A\rightarrow 0$. This value has been employed in the calculation of single-groove flow \citep{schoenecker2013}, yielding very accurate results. Additionally, one may want to take into account the distribution of the local slip length for shallow cavities, which further enhances the accuracy of the model at low $A$. 
In the lubrication limit, the local slip length is a constant. In contrast, within the current framework, the local slip length is modeled as an ellipse. As was shown by \citet{ybert2007}, in the limit of small slip, which is implied by $A\rightarrow 0$, the global behavior of the flow is determined only by the mean local slip length, independent of its distribution. Hence, to assure a consistent mean local slip in this limit, the normalized maximum local slip length of an ellical distribution has to fulfill
\begin{equation}
\lim\limits_{A\rightarrow 0}\lk.\frac{\dd D_\mr l}{\dd A}\rk|_{a=0}=\frac{4}{\pi}.
\label{eq:dabed2}
\end{equation}
The simplest model we can assume for the variation between these two limits is a linear dependency, i.e. 
\begin{equation}
\lim\limits_{A\rightarrow 0}\frac{\dd D_\mr l}{\dd A}=1\cdot g(a)\quad \text{with}\quad g(a)=\frac{4}{\pi}-\frac{4-\pi}{\pi}a.
\label{eq:dabeges}
\end{equation}
Despite the simplicity of this assumption, later it will become obvious that it captures the behavior at $A\rightarrow 0$ sufficiently well. 

The second coefficient now follows from eq. \eqref{eq:ansatzdlp}, such that the full parameterization of the longitudinal maximum local slip length is
\begin{equation}
D_\mr l(A,a)=f(a)\, \beta_{\mr{ l,\,N=1,\,max}}^*\,\erf\lk(\frac{g(a)\sqrt{\pi}}{2 f(a)\, \beta_{\mr{ l,\,N=1,\,max}}^*}\,A\rk).
\label{eq:dlp}
\end{equation}

\subsection{Transverse flow}

The maximum local slip length for transverse flow is obtained via a similar procedure. Again, the ansatz is
\begin{equation}
D_\mr t(A,a)=d_{0,\mr t}(a)\erf(d_{1,\mr t}(a)\,A).
\label{eq:ansatzdtp}
\end{equation}
Physically, the ansatz reflects the vortex pattern inside a groove or cavity. The increase of the error function with $A$ at low aspect ratios represents the growth of the primary vortex with the depth of the groove. Starting from around $A\approx 1$, gradually more vortices form at the bottom of the groove, while the position and size of the primary vortex remain essentially unaffected. Since for a purely apparent slip, the local slip can be expressed by the flow field close to the fluid-fluid interface, also the slip length saturates at high $A$, which is described by the plateau of the error function. This kind of vortex pattern is well known from lid-driven cavity flows (see \citet{pan1967,joseph1978,shankar1993}).

Since due to the more complex governing equations the analytical calculation of transverse flow is mathematically much more intricate than that of longitudinal flow, there are no analytical expressions for the behavior of the effective slip length with the fluidic interface fraction available. However, an analogy to the longitudinal flow may be drawn. 

It is assumed that for $A\rightarrow\infty$, $D_\mr t$ obeys the same behavior with respect to $a$ as $D_\mr l$, that is 
\begin{equation}
\lk.D_\mr t\right|_{A\rightarrow\infty}=f(a)\lk.D_\mr t\right|_{a=1,\,A\rightarrow\infty}=f(a) \,\beta_{\mr{ t,\,N=1,\,max}}^*.
\label{eq:Dta}
\end{equation}
The magnitude of the maximum effective slip length $\beta_{\mr{ t,\,N=1,\,max}}^*$ for $N=1$ can for example be obtained from the work of \citet{hocking1975}. He considered transverse shear flow over an array of infinitely thin parallel plates. In this limit, the effective slip length was found to be 
\begin{equation}
\beta_{\mr{ t,\,N=1,\,max}}^*=\frac{0.505}{2\pi}
\label{eq:hocking}
\end{equation}
for $N=1$. 

It is important to note that in Hocking's study indeed two different fluids with a flat fluid-fluid interface were considered. The resulting effective slip length is therefore somewhat smaller than that of a single-phase flow over the same geometry \citep{luchini1991, hocking1975}, where the the fluid streamlines may penetrate into the spaces between the plates. 
Hence, eq. \eqref{eq:hocking} is applicable to the present situation and we can write
\begin{equation}
d_{0,\mr t}(a)=f(a)\, \beta_{\mr{ t,\,N=1,\,max}}^*.
\label{eq:d0t}
\end{equation}

The second coefficient again follows from the limit $A\rightarrow 0$. 
If the dependence on $a$ is again analogous to the longitudinal case, lubrication theory leads to
\begin{equation}
\lim\limits_{A\rightarrow 0}\frac{\dd D_\mr t}{\dd A}=\frac{1}{4}\cdot g(a),
\label{eq:dabegest}
\end{equation}
and the expression for the maximum local slip length becomes 
\begin{equation}
D_\mr t(A,a)=f(a)\, \beta_{\mr{ t,\,N=1,\,max}}^*\,\erf\lk(\frac{g(a)\sqrt{\pi}}{8 f(a)\, \beta_{\mr{ t,\,N=1,\,max}}^*}\,A\rk).
\label{eq:dtp}
\end{equation}

Overall, the maximum local slip lengths \eqref{eq:dlp} and \eqref{eq:dtp} have been modeled heuristically. However, this description relies on physical arguments, not involving any numerical fitting procedures. In the limit of $a\rightarrow0$, the results for a periodically structured wall are consistent with those for a single groove \citep{schoenecker2013}. 

%%%%%%%%%%%%%%%%%%%%%%%%%%%%%%%%%%%%%%%%%%%%%%%%%%%%%%%%%%%%%%%%%
\section{Results and discussion}

With the mathematical expressions for the flow field and the effective slip length as well as the modeling of the maximum local slip length, the flow over an array of grooves is now fully described as a function of the groove aspect ratio, the fluidic interface fraction, and the viscosity ratio of the fluids flowing over and filling the grooves.

\subsection{Transverse flow}

\subsubsection{Flow field}

\begin{figure}
\centering
	\subfloat[][]{%
	\includegraphics[width=0.48\textwidth]{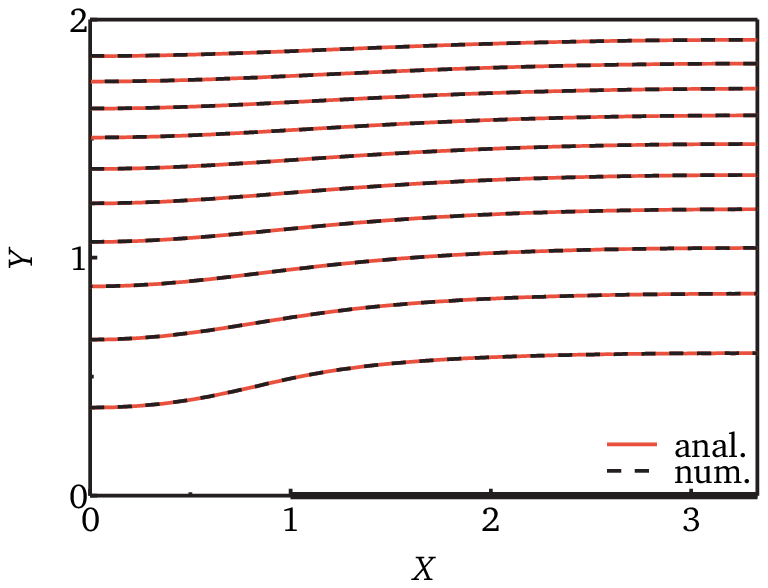}}
	\hspace{5pt}%
\subfloat[][]{%
	\includegraphics[width=0.48\textwidth]{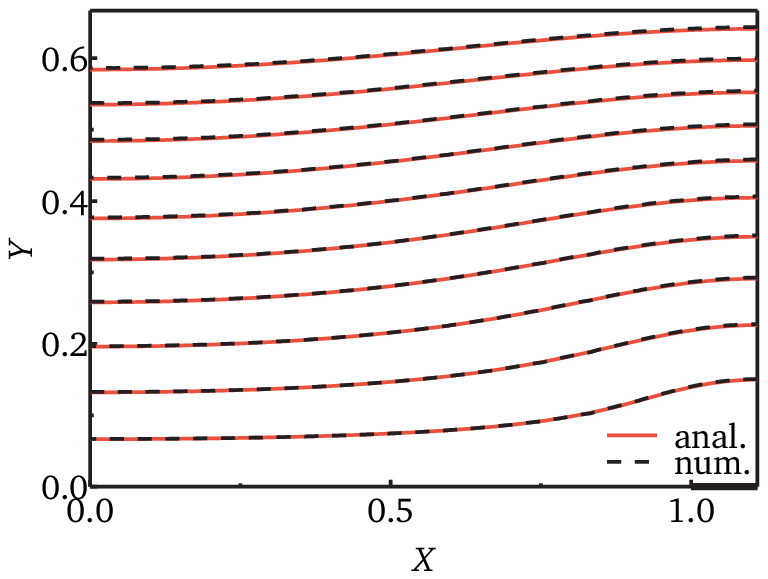}}\\
	\subfloat[][]{%
	\includegraphics[width=0.48\textwidth]{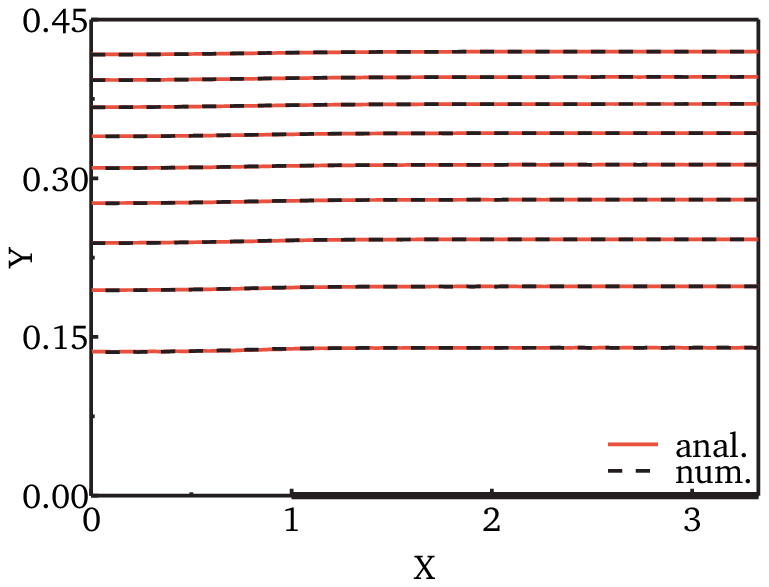}}
	\hspace{5pt}%
\subfloat[][]{%
	\includegraphics[width=0.48\textwidth]{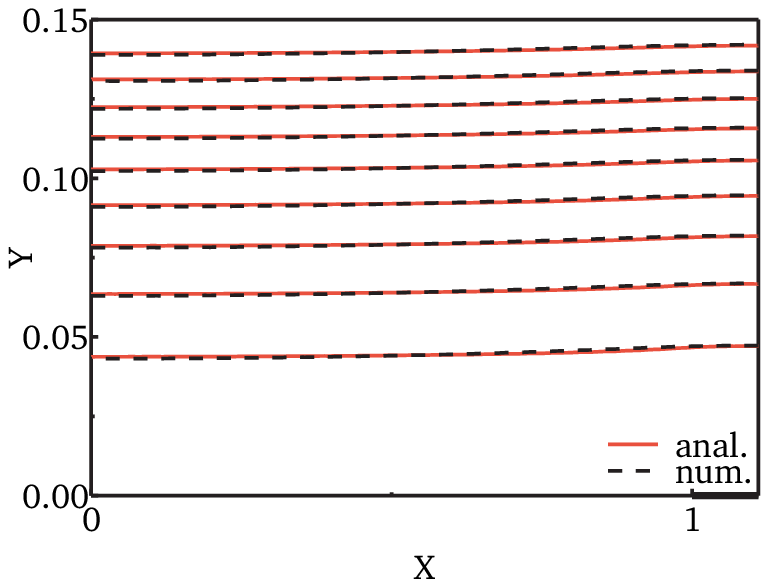}}	
	\caption{Normalized transverse streamline patterns (from eq. \eqref{eq:stromfkttransperioded} with \eqref{eq:dtp}) for $A=1$ and varying viscosity ratio and fluidic interface fraction: water flowing over air ($N\approx55.56$): a)~$a=0.3$; b)~$a=0.9$; air flowing over water ($N\approx0.02$): c)~$a=0.3$; d)~$a=0.9$. Numerically calculated streamlines are shown for comparison.
	}
	\label{fig:stromltransperiodavar}
\end{figure}

For transverse flow, examples of streamline patterns can be observed in fig. \ref{fig:stromltransperiodavar}. Subfigures (a) and (b) show the classical case of a superhydrophobic surface, where water ($\eta=10^{-3}$\,Pa\,s) flows over air-filled grooves ($\eta=1.8\times 10^{-5}$\,Pa\,s).
Such streamlines can be calculated for arbitrary fluidic interface fractions. The diagrams illustrate the change in the flow field, when $a$ is increased from $0.3$ to $0.9$. The opposite situation, air flowing over water-filled cavities, may arise through capillary condensation of water in surface grooves, which is shown in fig. \ref{fig:stromltransperiodavar} (c) and (d), again for fluidic interface fractions of $0.3$ and $0.9$. Indeed, the presence of water in the grooves changes the flow along the surface compared to a flat wall. However, this effect is relatively weak due to the small viscosity ratio of air and water. Therefore, it extends only over a region very close to the interface, as can be seen from the scales on the $y$ axis in fig. \ref{fig:stromltransperiodavar} (c) and (d). In all demonstrated cases, $A=1$.

\begin{figure}
\centering
	\subfloat[][]{%
	\includegraphics[width=0.48\textwidth]{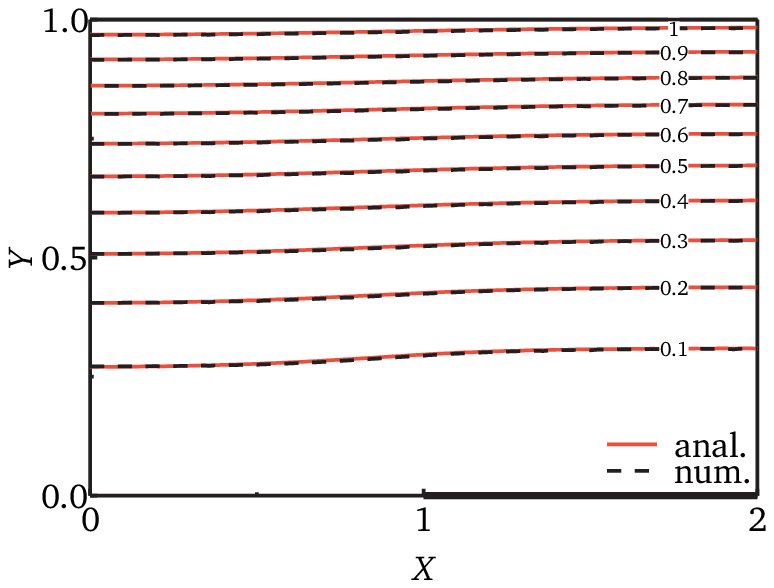}}
	\hspace{5pt}%
\subfloat[][]{%
	\includegraphics[width=0.48\textwidth]{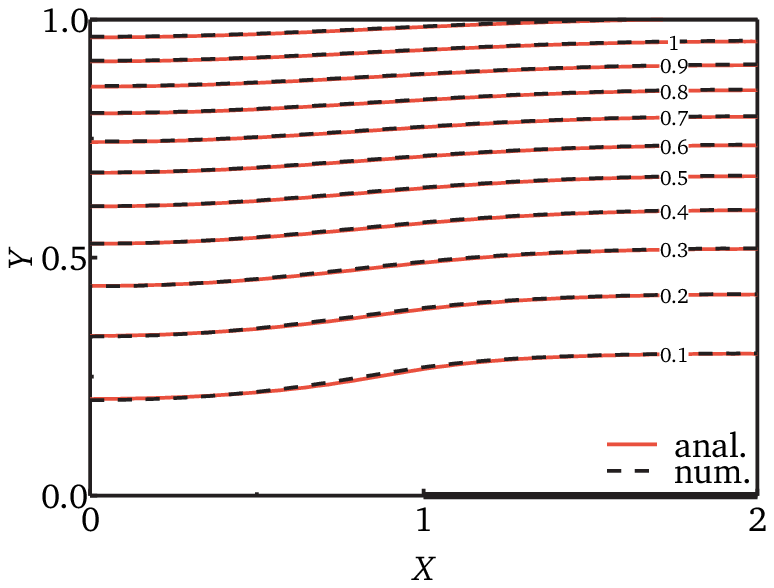}}
		\caption{Variation of the normalized transverse streamline patterns (from eq. \eqref{eq:stromfkttransperioded} with \eqref{eq:dtp}) for $N=1$ and $a=0.5$ with the groove aspect ratio: a)~$A=0.125$; b)~$A=2$. Numerically calculated streamlines are shown for comparison.}
	\label{fig:stromltransA}
\end{figure}

In order to verify the derived functional form of the flow field, numerical calculations have been performed with the commercial finite-element solver Comsol Multiphysics$^\text{\textregistered}$ for a range of parameters. In a periodic computatiional domain, which contains both the fluid above the surface and one groove with the second fluid, the Navier-Stokes equations have been solved for various viscosity ratios, aspect ratios and fluidic interface fractions. At a distance of 5$L$ from the surface, the flow was driven by a shear stress $\tau$. This is sufficiently far away from the surface for the flow field to be considered uniform. With $\tau=0.1$\,N/m$^2$ it was ensured that the Reynolds number $\Rey=\left.u\right|_{x=0,y=0} b \rho/\eta\ll1$ for both fluids. The triangular mesh was strongly refined along the fluid-fluid interface and the wall to adequately resolve the changes in the flow field due the different boundary and transition conditions, respectively. The number of mesh elements varied with the considered values of $a$ and $A$, starting from about 140000. With this, the numerical resolution is large enough to obtain virtually grid-independent results. After calculation of the flow field, the results were integrated to yield the stream function, whose isolines are also plotted in figs. \ref{fig:stromltransperiodavar} and \ref{fig:stromltransA}. 

In fig. \ref{fig:stromltransA}, the development of the flow field with increasing depth of the grooves is illustrated. With the aspect ratio $A$ rising from 0.125 (fig. \ref{fig:stromltransA} (a)) to 2 (fig. \ref{fig:stromltransA} (b)), the streamlines are notably drawn closer to the surface, indicating an increase in velocity. This behavior is influenced by the viscosity ratio. The examples of fig. \ref{fig:stromltransA} correspond to a viscosity ratio of $N=1$. For the fluids considered in fig. \ref{fig:stromltransperiodavar}, the flow field is much less sensitive to the change in  aspect ratio. In the case of water flowing over air, the reduction in flow resistance is already so high at $A=0.125$ that the further change when $A=2$ is relatively small. In the opposite case, the relative change is bigger, although the absolute influence on the flow is very small due to the small value of $N$. This fact can also be understood from the equation for the stream function \eqref{eq:stromfkttransperioded}. $N$ and $A$ always enter the equation in combination as a part of the net local slip \eqref{eq:nettoslip}. The order of magnitude of the slip contribution in \eqref{eq:stromfkttransperioded} is $\sim N D_\mr t/(1+N D_\mr t)$ with $D_\mr t$ increasing monotonically with $A$. $D_\mr t$ is always of the order of $10^{-1}$, while $N$ can span a wide range of orders of magnitude. If the magnitude of $N$ is much larger than that of $D_\mr t$, small changes in $D_\mr t$ will not affect the dominance of the slip contribution to the flow field.

In general, the mathematical structure of eq. \eqref{eq:stromfkttransperioded} exhibits the parametric dependencies quite well. The full solution is a superposition of a Couette flow and Philip's classical solution, with the Couette flow contribution being $\sim 1/(1+C_\mr t)$, and the contribution of Philip's solution being $\sim C_\mr t/(1+C_\mr t)$. Depending on the order of magnitude of $C_\mr t$, one or the other term dominates. The parameter $C_\mr t$ is of the order of $N D_\mr t$ for basically all values of $a$. Then, the magnitude of $N D_\mr t$ relative to 1 determines the behavior of the flow field. The same dependency applies to the effective slip length, as can be observed from eq. \eqref{eq:betatedc}.

The agreement between the analytically and numerically calculated streamlines is excellent. Fig. \ref{fig:stromltransperiodavar} and \ref{fig:stromltransA} virtually span the whole range of viscosity ratios, demonstrating a good agreement everywhere. This is even true in close proximity to the fluid-fluid interface (fig. \ref{fig:stromltransperiodavar} (c)). The variation of the flow field with the $A$ is also captured well, as demonstrated for $A=0.125$ and $A=2$ in fig.~\ref{fig:stromltransA} and which is also true for more extreme or mid-sized values of $A$. Any deviations between the analytical and numerical results can only be observed for low viscosity ratios in extreme proximity to the interface with respect to the groove length scale. 
This is due to the behavior at the fluid-fluid interface itself, where the local slip-length distribution of the model may not fully match the actual distribution.

\begin{figure}
\centering
	\subfloat[][]{%
	\includegraphics[width=0.48\textwidth]{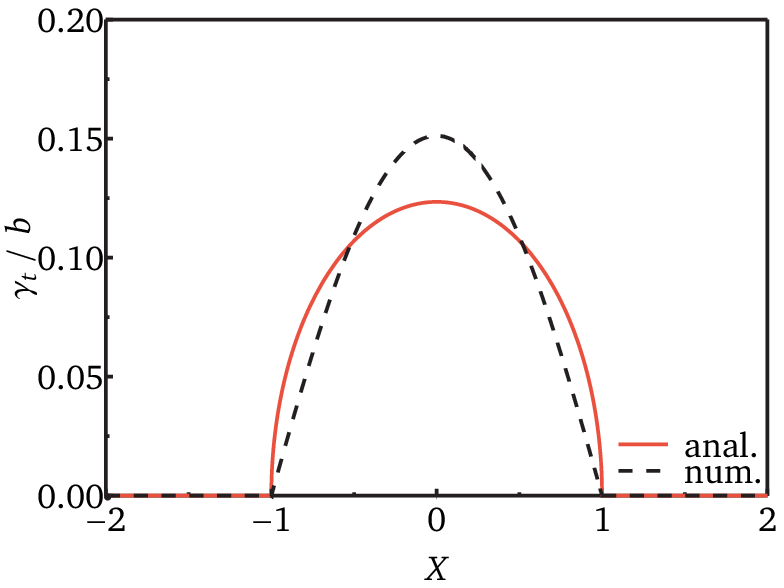}}
	\hspace{5pt}%
\subfloat[][]{%
	\includegraphics[width=0.48\textwidth]{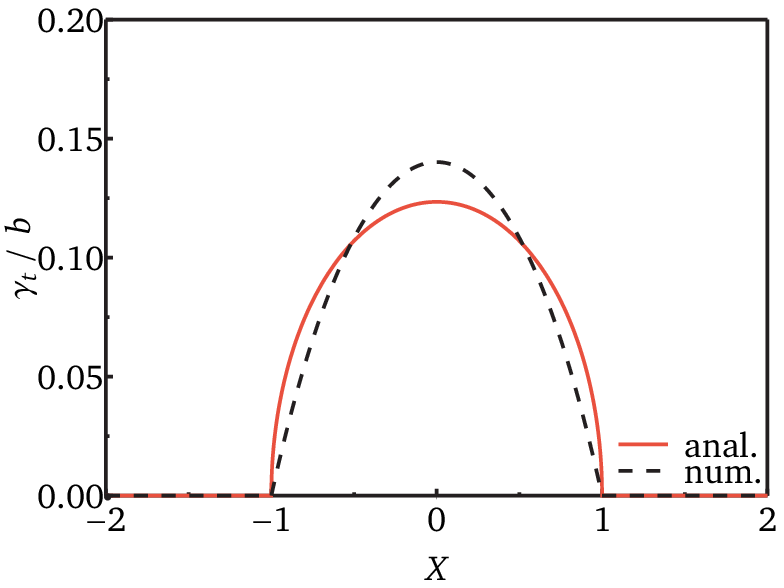}}\\
\subfloat[][]{%
	\includegraphics[width=0.48\textwidth]{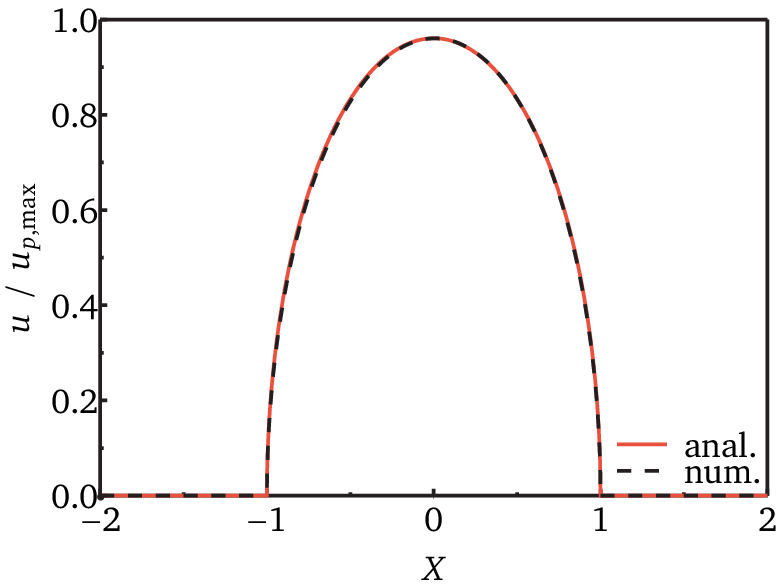}}
	\hspace{5pt}%
\subfloat[][]{%
	\includegraphics[width=0.48\textwidth]{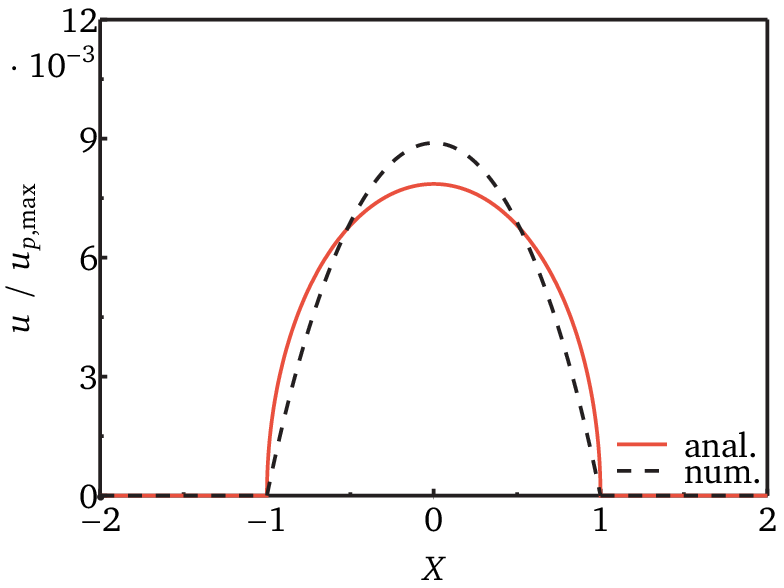}}	
	\caption{Comparison of the analytical and numerical results at $Y=0$ for $A=1$ and $a=0.5$: normalized slip length distribution (\eqref{eq:gammat} with magnitude \eqref{eq:dtp}) for: a)~water flowing over air-filled grooves ($N\approx 55.56$) and b)~air flowing over water-filled grooves ($N\approx 0.02$); normalized velocity (from eq. \eqref{eq:stromfkttransperioded}) for: c)~water on air and d)~air on water.
	}
	\label{fig:transrandvgl}
\end{figure}

The situation directly at the interface between the two fluids at $Y=0$ is shown in fig. \eqref{fig:transrandvgl}. For the two familiar examples of water flowing over air-filled grooves and air flowing over water-filled grooves, a comparison of the analytically and numerically calculated local slip length distributions is given in subfigures (a) and (b), and of the corresponding velocities in subfigures (c) and (d). The velocities are normalized with the maximum velocity at $X=Y=0$ in Philip's case for the same fluidic interface fraction $u_{\mr{p,\,max}}=\frac{\tau_\infty}{\eta_1}\frac{L}{2 \pi}\imag(\arccos(\sec(\alpha)))$. The results are very similar to those for a single groove \citep{schoenecker2013}. The agreement between the analytically computed local slip length (eq. \eqref{eq:gammat} with \eqref{eq:dtp}) and the  numerical value $u_1/(N \pd u_1/\pd y)$ is reasonably good. Considering that the choice of the the local slip distribution in the model is mathematically motivated, the agreement is quite remarkable. The similarity of the numerical results in fig. \eqref{fig:transrandvgl} (a) and (b) for viscosity ratios as different as in the chosen examples supports one of the main assumptions of the model, namely the existence of a universal local slip length distribution, which is characteristic to the geometry of the surface and independent of the involved fluids. 

As discussed, the magnitude of the term $N D_\mr t$ relative to 1 is responsible for the character of the flow field and hence for the accuracy of the model. Since the contribution of Philip's solution to the flow field behaves as $\sim N D_\mr t/(1+N D_\mr t)$ for practically all values of $a$, a large value for $N$ makes the overall flow insensitive to changes in the local slip length as long as the magnitude of $N D_\mr t$ is much larger than 1 (subfigure (a) vs. (c)). This also explains why other studies of slip at conventional superhydrophobic surfaces obtained reasonable results despite employing different local slip-length distributions or magnitudes \citep[see e.g.][]{mongruel2013}. 
Furthermore, it is interesting to note that the integral values of the analytically and numerically calculated slip lengths and velocities are always consistent, which assures a good accuracy of the far-field behavior of the flow field as well as of the effective slip length. 
As the effective slip length is a far-field effect, or alternatively interpreted, an effect only determined the mean velocity at $Y=0$ (cf. eq. \eqref{eq:sliptgemittelt}), it is expected that also this quantity is accurately predicted. 

Overall, the expression for the stream function \eqref{eq:stromfkttransperioded} with local slip length \eqref{eq:gammat} of magnitude \eqref{eq:dtp} is excellently suited to represent the effect of rectangular grooves on the transverse flow above them, for a fluid of any viscosity filling the grooves, for any surface coverage of the grooves, as well as any aspect ratio. The analytical results are accurate as well as simple at the same time.

\subsubsection{Effective slip length}

\begin{figure}
\centering
	\subfloat[][]{%
	\includegraphics[width=0.48\textwidth]{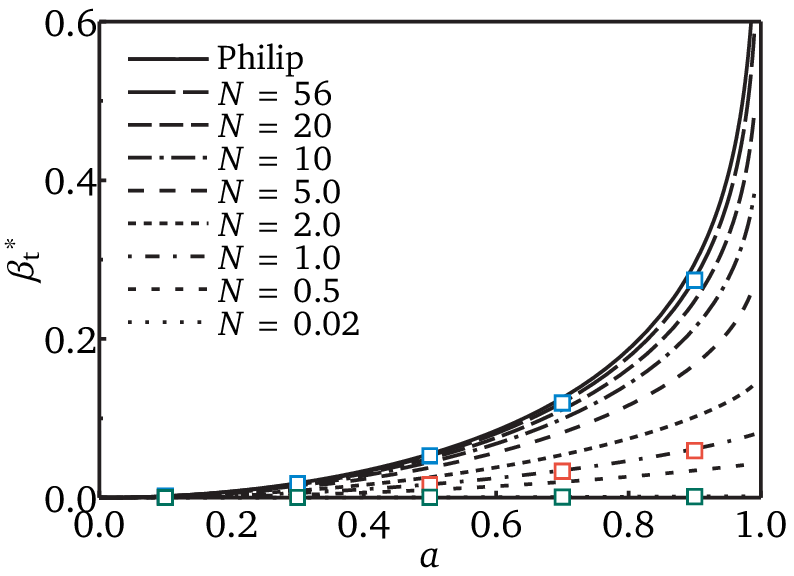}}
	\hspace{5pt}%
\subfloat[][]{%
	\includegraphics[width=0.48\textwidth]{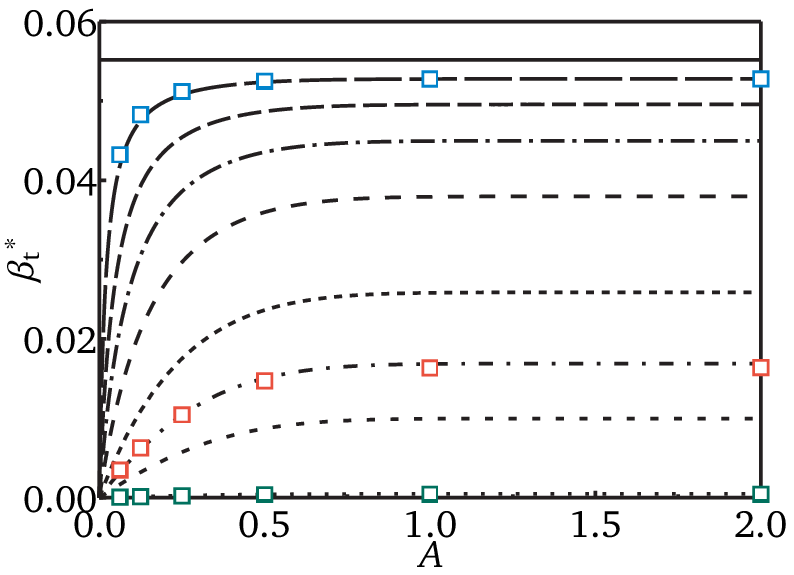}}
	\caption{Transverse effective slip length (\eqref{eq:betated} with \eqref{eq:dtp}) for various viscosity ratios: (a)~as a function of the fluidic interface fraction at $A=1$; (b)~as a function of the aspect ratio at $a=0.5$. Squares indicate numerically calculated values. }
	\label{fig:effsliptrans}
\end{figure}

The transverse effective slip length has also been investigated for different parameter combinations. Fig. \ref{fig:effsliptrans}~(a) shows the evolution of $\beta_\mr t^*$ with the fluidic interface fraction $a$ at an exemplary value of $A=1$. The differently dashed lines correspond to a variety of viscosity ratios, where $N=56$ stands for the special case of water flowing over air-filled grooves and $N=0.02$ for the reverse case. While Philip's solution for perfectly slipping stripes tends to infinity as $a\rightarrow 1$, eq. \eqref{eq:betated} leads to finite values, depending on the respective viscosity ratio of the two fluids. For the same values of $N$, fig. \ref{fig:effsliptrans}~(b) shows the variation of $\beta_\mr t^*$ with the aspect ratio $A$ at $a=0.5$. With increasing $A$, $\beta_\mr t^*$ first also increases and then saturates at its final value for the respective viscosity ratio $N$. The higher the viscosity ratio, the earlier this final value is achieved. Superhydrophobic surfaces, as with $N=56$, obviously do not need to exhibit deep grooves to provide a considerable slip. Already at $A=0.25$, the maximum value is virtually reached. This is interesting from the viewpoint of surface design. Surface structures with an aspect ratio greater than 1 are usually hard to manufacture, so it is advantageous if they can be avoided. There might be other reasons to choose an aspect ratio greater than 0.25, like the prevention of a Cassie-to-Wenzel transition, but in terms of slip, no significant further gain can be achieved. 

To compare with the analytical model, the effective slip length was also extracted from the numerical calculations described above using eq. \eqref{eq:sliptgemittelt}. The agreement between the analytical and numerical results is very good over the complete range of viscosity ratios and aspect ratios, as can be observed 
from fig. \ref{fig:effsliptrans}.

\subsection{Longitudinal flow}

\begin{figure}
\centering
	\subfloat[][]{%
	\includegraphics[width=0.48\textwidth]{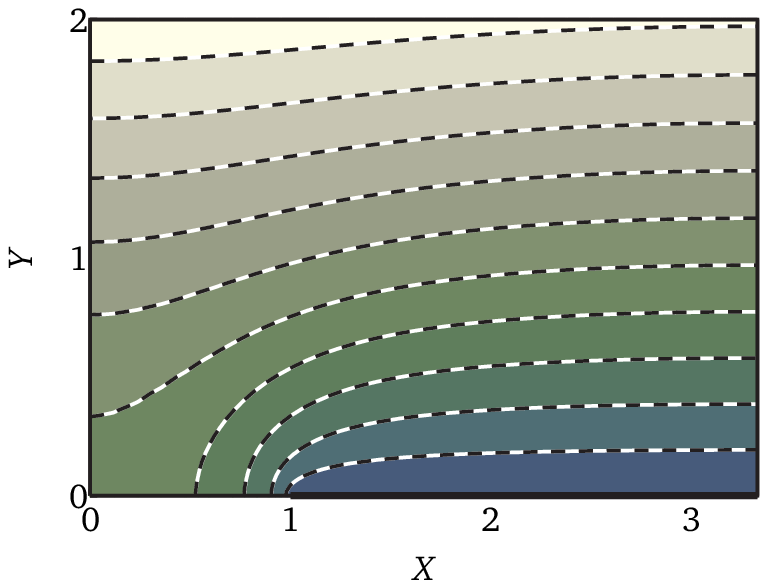}}
	\hspace{5pt}%
\subfloat[][]{%
	\includegraphics[width=0.48\textwidth]{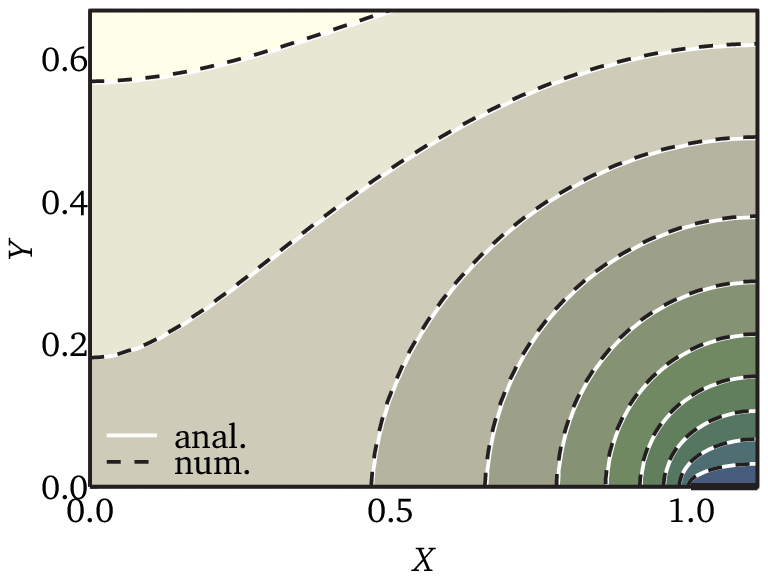}}\\
\subfloat[][]{%
	\includegraphics[width=0.48\textwidth]{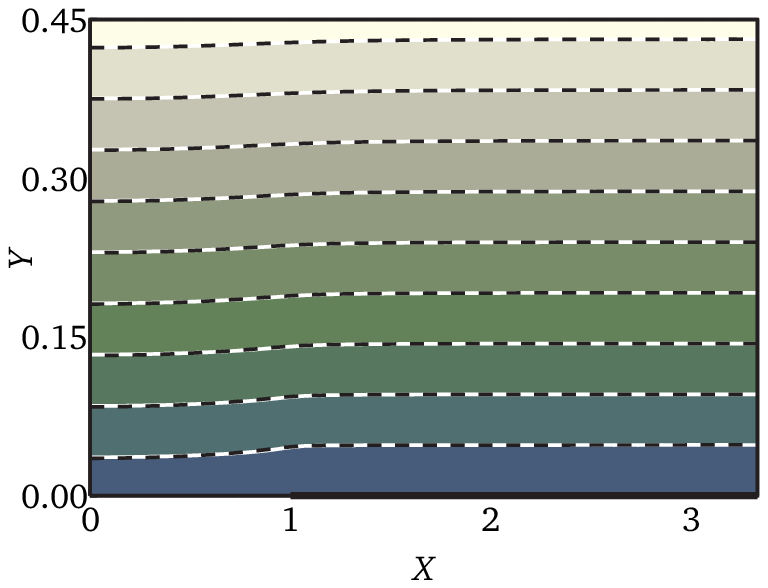}}
	\hspace{5pt}%
\subfloat[][]{%
	\includegraphics[width=0.48\textwidth]{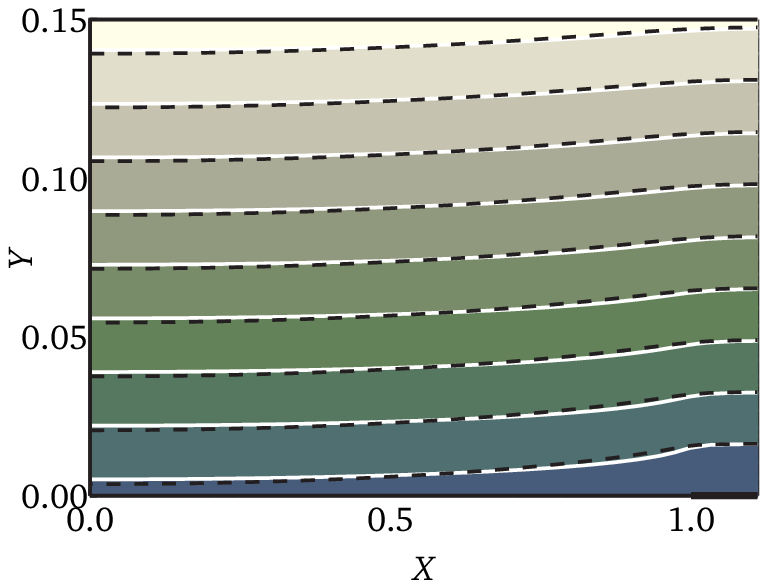}}	
	\caption{Normalized longitudinal velocity $\tilde w$ (eq. \eqref{eq:wlongperioded} with \eqref{eq:dlp}) for $A=1$ and varying viscosity ratio and fluidic interface fraction: water flowing over air ($N\approx55.56$) for: a)~$a=0.3$ and b)~$a=0.9$; air flowing over water ($N\approx0.02$) for: c)~$a=0.3$ and d)~$a=0.9$. Numerically calculated velocity isolines are shown as dashed lines for comparison.
	}
	\label{fig:ulongperiodavar}
\end{figure}

For longitudinal flow over a periodic array of grooves, the variation of the velocity $\tilde w$ with the fluidic interface fraction is depicted in fig. \ref{fig:ulongperiodavar}, again for a conventional superhydrophobic surface (subfigures (a) and (b)) and its reverse configuration(subfigures (c) and (d)). Numerical solutions of the Laplace equation were again computed with Comsol Multiphysics$^\text{\textregistered}$ and are shown with dashed lines in the same figure. All other conditions, e.g. computational domain and mesh, are analogous to the transverse configuration. The agreement between analytical and numerical results is very good. At $Y=0$ and in very close vicinity to this line, small local deviations but integral agreement can be observed, similar to the transverse case. Note that in subfigure (d) the $Y$ axis is scaled to a region very close to the interface. The variation of the flow field with the aspect ratio of the grooves is also analogous to the transverse case. In total, a longitudinal flow over a periodic array of rectangular grooves containing a second immiscible fluid is well characterized by eq. \eqref{eq:wlongperioded} and \eqref{eq:dlp} over the whole range of parameters. 

\begin{figure}
\centering
	\subfloat[][]{%
	\includegraphics[width=0.48\textwidth]{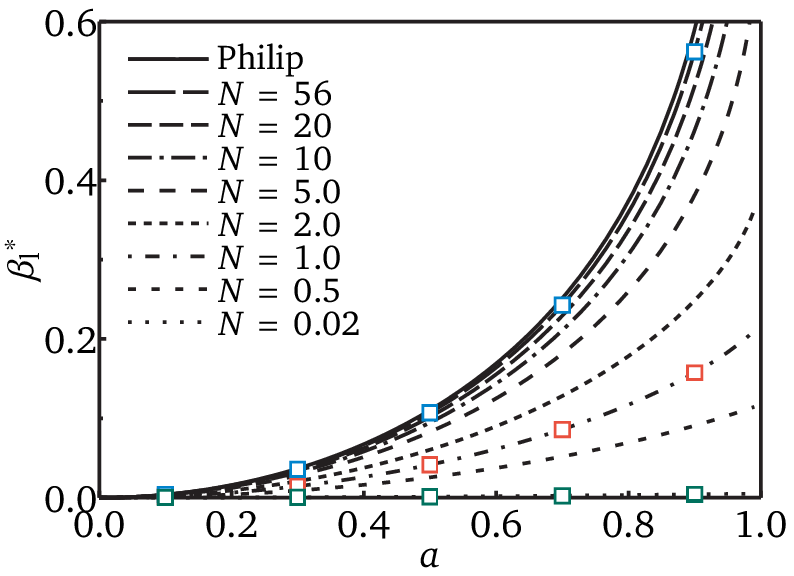}}
	\hspace{5pt}%
\subfloat[][]{%
	\includegraphics[width=0.48\textwidth]{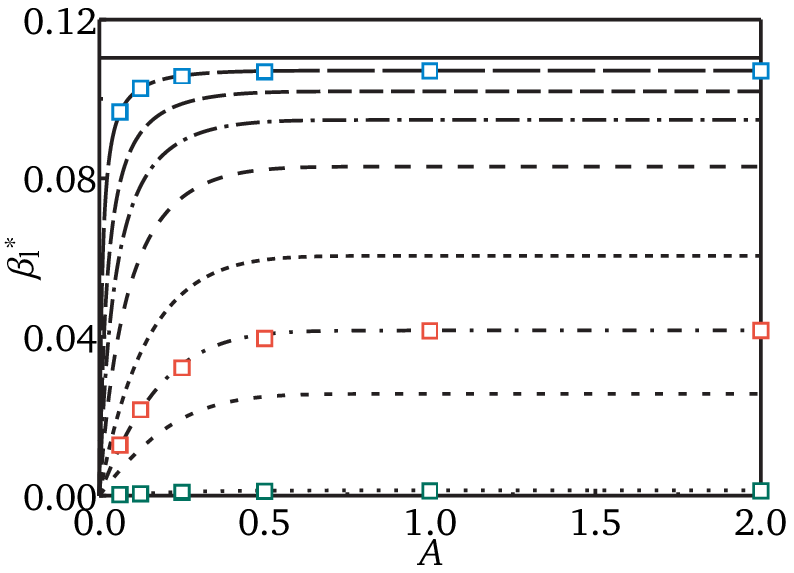}}
	\caption{Longitudinal effective slip length (eq. \eqref{eq:betaledc} with \eqref{eq:dlp}) for various viscosity ratios: (a)~as a function of the fluidic interface fraction at $A=1$; (b)~as a function of the aspect ratio at $a=0.5$. Squares indicate numerically calculated results. }
	\label{fig:effsliplong}
\end{figure}

This fact is also obvious from the results for the longitudinal effective slip length (fig. \ref{fig:effsliplong}). Consistently, analytical and numerical results agree very well. Compared to the transverse slip length, the longitudinal effective slip length is generally slightly more than twice as large. An exact factor of two occurs in the case of a perfectly slipping fluid-fluid interface. The general behavior of the longitudinal effective slip length with varying fluidic interface fraction and aspect ratio is similar to the transverse case.

Finally, it should be noted that for $N=1$ the equations for longitudinal flow  also apply to single-phase flow, e.g. the flow of air or water along an obstacle. This is not the case for transverse flow, where the streamlines may cross the line $y=0$ in single-phase flow.

%%%%%%%%%%%%%%%%%%%%%%%%%%%%%%%%%%%%%%%%%%%%%%%%%%%%%%%%%%%%%%%%%
%%%%%%%%%%%%%%%%%%%%%%%%%%%%%%%%%%%%%%%%%%%%%%%%%%%%%%%%%%%%%%%%%
%\section{Consequences} 
\subsection{Relation to existing solutions} 

To the best of the authors' knowledge, for the first time expressions for the flow field and the effective slip length have been derived specifically containing the viscosities of the involved fluids and information about the shape of the surface features. Despite this complexity, the equations are closed analytical expressions. Compared to generally employed Fourier series expansions, where infinite sums have to be evaluated (e.g. \citet{wang2003}), this has great advantages in the handling of the equations. Furthermore, the influence of the various parameters, like the viscosity ratio or the fluidic interface fraction, is directly exhibited. At the same time, the equations are very accurate, so that the effective slip length of a surface with rectangular grooves can be calculated extremely precisely over the whole range of parameters. This allows the effective slip length to be used in numerical simulations of multiscale problems, where it strongly reduces the computational complexity while giving the opportunity of performing parameter variations, e.g. in optimization problems. 
The modeling of the local slip length is based on physical considerations. There are no numerical fitting procedures or even tunable parameters involved.

%%%%%%%%%%%%%%%%%%%%%%%%%%%%%%%%%%%%%%%%%%%%%%%%%%%%%%%%%%%%%%%%%%%%%%%%%5

In the limit of an inviscid fluid in the grooves, i.e. $N\rightarrow \infty$, the equations for the flow field \eqref{eq:stromfkttransperioded} and \eqref{eq:wlongperioded} as well as those for the effective slip length \eqref{eq:betated} and \eqref{eq:betaledc} reduce to the known expressions of \citet{philip1972} for an array of no-shear slots in a no-slip wall. This case corresponds to an infinite net local slip length $\Gamma_\mr{net, local}=N \gamma$, 
and is consistent.

Besides \citeauthor{philip1972}'s results, it is interesting to compare the equations for the effective slip length \eqref{eq:betated} and \eqref{eq:betaledc} to the 
equations of \citet{belyaev2010} \eqref{eq:betabvt} and \eqref{eq:betabvl}. In this work, the authors aim at describing a superhydrophobic surface with a constant local slip length at the air-water interfaces. This slip length, which is termed $b_{\mr{BV}}$ here, is of unspecified magnitude. 

Remarkably, when choosing the local slip of \citet{belyaev2010} to be $b_{\mr{BV}}=N d_\mr t$ and $N d_\mr l$, respectively, equations \eqref{eq:betabvt} and \eqref{eq:betabvl} exactly coincide with the present effective slip equations \eqref{eq:betated} and \eqref{eq:betaledc}. This value for $b_{\mr{BV}}$ is just the maximum of the net local slip of the  model presented in this article.

Several consequences can be deduced from this relationship. Firstly, the results of \citet{belyaev2010} are in principle confirmed, since by appropriately selecting $b_{\mr{BV}}$ they can be rendered equivalent to equations \eqref{eq:betated} and \eqref{eq:betaledc}, which were obtained on a completely different way. Secondly however, the connection between both equations sheds light on the significance of the simplifications made by \citet{belyaev2010}. Starting from a constant local slip distribution, a certain term in their eq. (2.19) was discarded in the course of the calculation. Obviously, the neglected term just represents the difference between a constant and a curved local slip distribution, both of maximum value $b_{\mr{BV}}$. Thus, the equations of \citet{belyaev2010} in fact correspond to a local slip length of the form of $\gamma_\mr{t/l}(x)$, given in eqs. \eqref{eq:gammat} and \eqref{eq:gammal}, with maximum value $b_{\mr{BV}}$ instead of a constant local slip of the same magnitude. This fact also explains why \citet{belyaev2010} observe a slight underestimation of the effective slip length when comparing their results to numerical calculations based on a constant local slip length.

%%%%%%%%%%%%%%%%%%%%%%%%%%%%%%%%%%%%%%%%%%%%%%%%%%%%%%%%%%%%%%%%%%%%%%%%%%%%%%
\section{On the ideality of a superhydrophobic surface}

As already introduced at the beginning of this article, the air-water interface of a superhydrophobic surface has often been modeled as a no-shear condition. 
This assumption suggests itself, since the viscosity of the enclosed air is very low. However, up to now its accuracy has not been verified. 

\begin{figure}
\centering
	\subfloat[][]{%
	\includegraphics[width=0.48\textwidth]{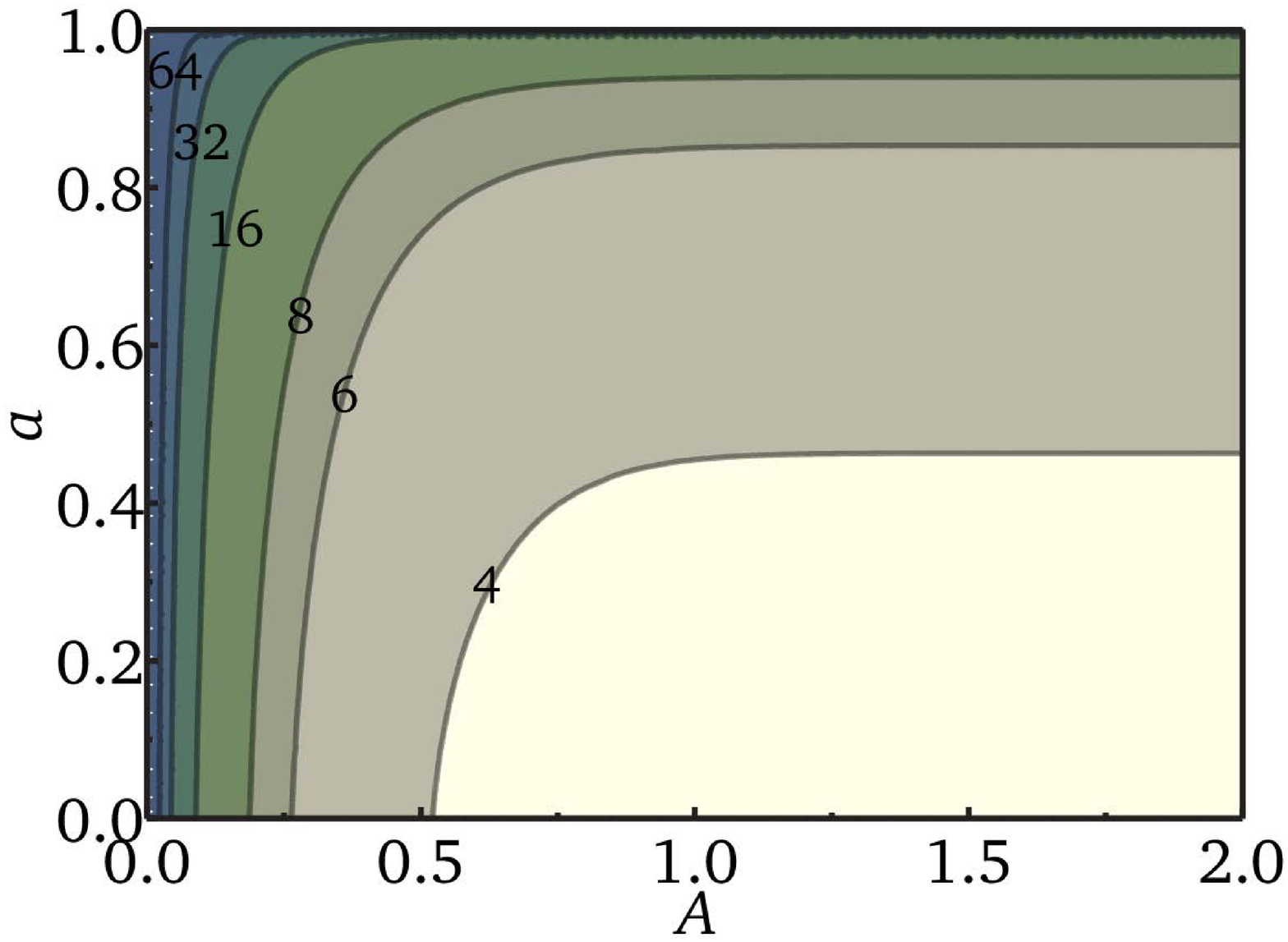}}
	\hspace{5pt}%
\subfloat[][]{%
	\includegraphics[width=0.48\textwidth]{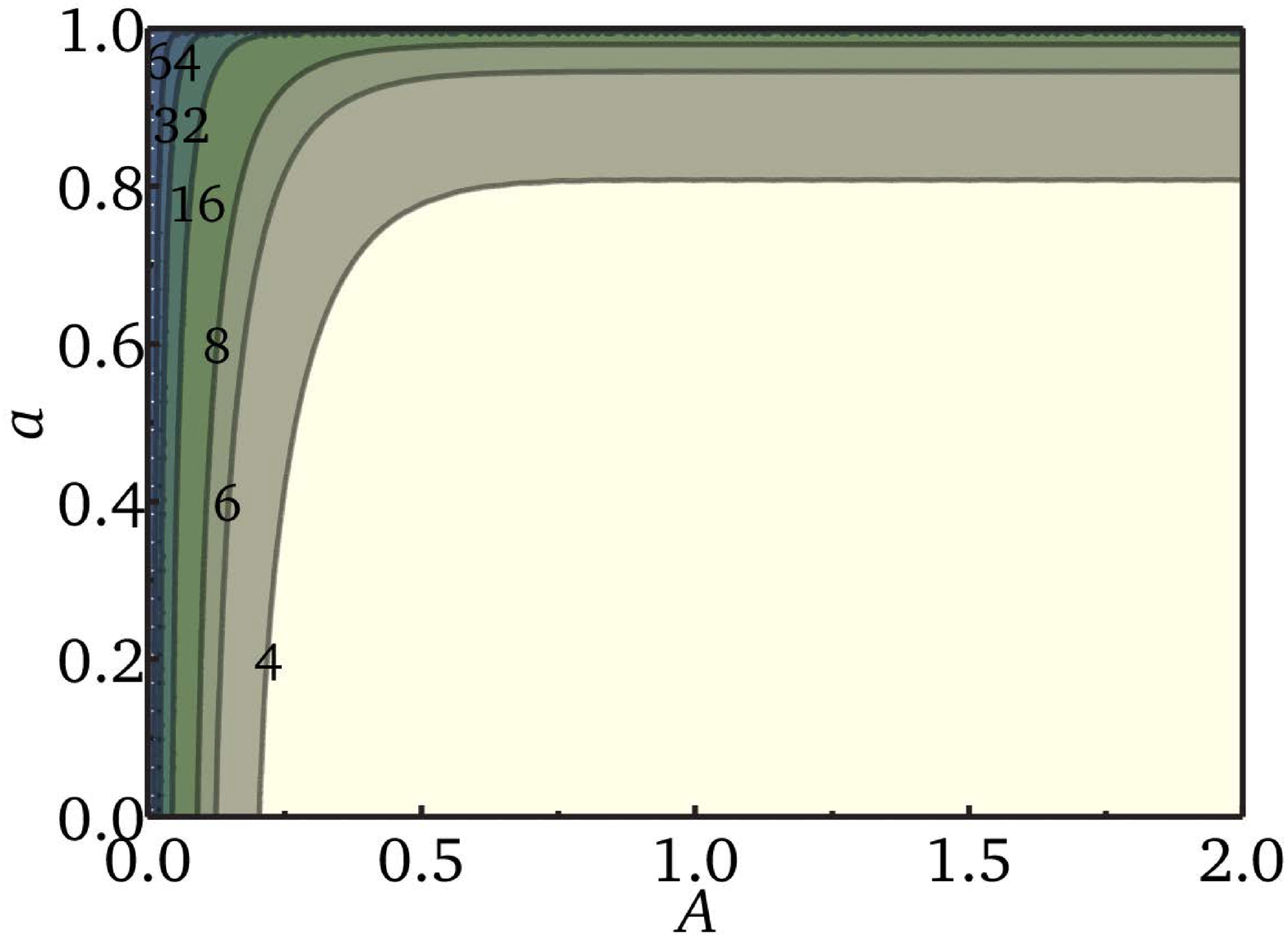}}
		\caption{Overestimation of the effective slip length in percent, if the water-air interface is assumed to be perfectly slipping: (a) transverse case, (b) longitudinal case}%
\label{fig:errormap}%
\end{figure}

The equations for the effective slip length \eqref{eq:betated} and \eqref{eq:betaledc}, together with the magnitude of the local slip length \eqref{eq:dtp} and \eqref{eq:dlp}, now allow investigating the accuracy of the no-shear assumption. In fig. \ref{fig:errormap}, the relative error in the effective slip length, when the air-water interface of a surface with rectangular grooves is assumed to be perfectly slipping $(\beta^*_\mr P-\beta^*)/\beta^*$ is shown in percent, both for transverse and longitudinal flow. $\beta^*_\mr P$ is Philip's solution for the normalized effective slip length.

Naturally, the no-shear assumption leads to an overestimation of the effective slip length. For large aspect ratios of the grooves, i.e. deep grooves, and a low surface coverage of the slipping areas, this overestimation is moderate. It stays within reasonable limits for a significant range of $A$ and $a$, with a better agreement in the longitudinal case. The minimum deviation between the effective slip of a superhydrophobic surface and Philip's surface is about 3.6\,\% for the transverse case and 2.6\,\% for the longitudinal case. 

Still, towards low aspect ratios as well as at high fluidic interface fractions, the error increases rapidly. At low aspect ratios, film flow occurs inside the grooves. Then, the gradient of the water velocity $\pd u_1/\pd y$ at the interface gets influenced by the thickness of the film. At high fluidic interface fractions, Philip's effective slip length tends to infinity, which is obviously not the case for any real fluid filling the grooves. This strong growth of $\beta^*_\mr P$ with $a$ apparently leads to a rapidly increasing overestimation for fluidic interface fractions close to 1. At the same time, fluidic interface fractions close to 1 are the technically most relevant. Therefore, we can conclude that the classical model for slip on superhydrophobic surfaces is inaccurate especially for one of the key scenarios currently studied.

%%%%%%%%%%%%%%%%%%%%%%%%%%%%%%%%%%%%%%%%%%%%%%%%%%%%%%%%%%%%%%%%%%%%%%%%%%%%%%%%%%%%
\section{A mathematical relation for the facilitated calculation of flow over slipping surfaces and the anisotropy of the local slip length}

The structure of the mathematical description of the flow field and the effective slip length in section \ref{matheperiodisch} already indicates a formal relation between transverse and longitudinal flow. 
In fact, a relation between transversal and longitudinal flow directly follows from the Goursat theorem.

In their entirety, the boundary conditions \eqref{eq:bcreel1p} and \eqref{eq:bcreel2p} for transverse flow and \eqref{eq:bcreel1lp} and \eqref{eq:bcreel2lp} for longitudinal flow over an array of grooves read in terms of Goursat functions
\begin{align}
\imag(W_\mr{t})&=2 N \gamma_\mr{t}(x)\reel(W'_\mr{t}) &\for &0\leq x \leq\frac{b}{2},\quad y=0  \label{eq:bcimag1p}\\
\imag(W_\mr{t})&=0 &\for &\frac{b}{2}< x \leq\frac{L}{2},\quad y=0  \label{eq:bcimag2p}\\
\reel(W'_\mr{t})&= \frac{\tau_\infty}{4\eta_1} &\for & y \rightarrow \infty
\end{align}
and
\begin{align}
\imag(W_\mr l)&=N \gamma_\mr{l}(x)\reel(W'_\mr l) &\for &0\leq x \leq\frac{b}{2},\quad y=0  \label{eq:bcimag1lp}\\
\imag(W_\mr l)&=0 &\for &\frac{b}{2}< x \leq\frac{L}{2},\quad y=0  \label{eq:bcimag2lp}\\
\reel(W'_\mr l)&= \frac{\tau_\infty}{\eta_1} &\for &y \rightarrow \infty.
\end{align}
Both $W_\mr t$ and $W_\mr l$ are analytic functions, hence they obey the Laplace equation and the solution is fully defined by the boundary conditions.

From the above equations, it is obvious that if there existed a longitudinal flow $\hat w=\imag(\hat W_\mr l)$ with a local slip length of $\hat \gamma_\mr l (x)=2 \gamma_\mr{t}(x)$, which is driven by a shear stress of $\hat\tau_\infty=\tau_\infty /4$, then the transverse and longitudinal Goursat functions would be equal. The transverse flow is thus mathematically connected to a longitudinal flow of double local slip and one fourth of the driving shear stress. Since the system is linear, the Goursat theorem \eqref{eq:Goursat} gives
\begin{equation}\label{eq:translongstroemung}
\begin{split}
\Psi_\mr t&=2 y \imag (\hat W_l)=2 y \frac{1}{4}\imag \lk(\lk.W_l\rk|_{\gamma_\mr l(x)=2 \gamma_\mr{t}(x)}\rk)\\
&=\frac{1}{2}y \lk. w\rk|_{\gamma_\mr l(x)=2 \gamma_\mr{t}(x)}.
\end{split}
\end{equation}
The notation $\lk. w\rk|_{\gamma_\mr l(x)=2 \gamma_\mr{t}(x)}$ shall indicate that the longitudinal velocity $w$ is driven by a shear stress of the same magnitude as the transverse flow, i.e. $\tau_\infty$, but the local slip is substituted as indicated. The velocity components of the transverse flow can be calculated from the stream function via the usual relations. Specifically, at $y=0$
\begin{equation}
\lk.u\rk|_{y=0}=\frac{1}{2}\lk.w\rk|_{\gamma_\mr l(x)=2 \gamma_\mr{t}(x),\ y=0}.
\label{eq:translongsurface}
\end{equation}
The corresponding evaluation of the effective slip length \eqref{eq:sliptgemittelt} yields
\begin{equation}
\beta_\mr t=\frac{1}{2}\lk.\beta_l\rk|_{\gamma_\mr l(x)=2 \gamma_\mr{t}(x)}.
\label{eq:translongbeta}
\end{equation}
Hence, the transverse effective slip length is equal to one half of the longitudinal effective slip length corresponding to a local slip distribution, which is twice the transverse one. 

Despite illustrated here using the boundary conditions representing a periodic array of grooves, equations \eqref{eq:translongstroemung}, \eqref{eq:translongsurface} and \eqref{eq:translongbeta} are valid for an arbitrary local slip distribution as long as the flow field is two-dimensional. 
The argumentation is merely based on Goursat's theorem, which is not restricted to any specific kind of boundary condition. 

The mathematical form of the relations \eqref{eq:translongstroemung}, \eqref{eq:translongsurface} and \eqref{eq:translongbeta} is the same as already derived by \citet{asmolov2012}. While their results were obtained through Fourier series expansions, here the relations are based on Goursat's theorem and confirm \citeauthor{asmolov2012}'s equations.

However, there is one significant difference in the basic assumptions leading to the above relations. \citeauthor{asmolov2012}'s derivation is based on a surface that exhibits the same local slip in both transverse and longitudinal direction. 
In this case, their equations corresponding to \eqref{eq:translongstroemung} to \eqref{eq:translongbeta} relate transverse and longitudinal flow over the same surface.

In fact, it has been shown in this article, for example in the equations for $D_\mr t$ \eqref{eq:dtp} and $D_\mr l$ \eqref{eq:dlp}, that for a grooved surface, the transverse and longitudinal local slip lengths are different. Hence, for such surfaces, the local slip length itself is anisotropic.

It is thus important to understand that in general the relations \eqref{eq:translongstroemung} to \eqref{eq:translongbeta} do not describe a connection between transverse and longitudinal flow over the same surface, but relate the transverse flow to a longitudinal flow over another hypothetical surface. 
It may therefore not be sufficient to only consider longitudinal flow along a surface and to deduce the transverse case form this solution, as sometimes suggested in the literature. The relations \eqref{eq:translongstroemung} to \eqref{eq:translongbeta} are in fact a method for a convenient calculation of the velocity field and slip length when the two-dimensional flow is governed by the biharmonic equation. They can reduce the mathematical effort significantly, since a solution of the Laplace equation is much less complicated than a solution of the biharmonic equation.

An anisotropic local slip also has consequences on the modeling of flow in various directions along patterned surfaces. Several authors have worked on mobility tensors, representing the effective slip length in arbitrary directions along a surface patterned with areas of local slip \citep{kamrin2010, kamrin2011, zhou2012, schmieschek2012, asmolov2012c, asmolov2012b} or on the effective slip length in different directions along patterned surfaces \citep{belyaev2010, feuillebois2010, mongruel2013}. In all of these investigations, an isotropic local slip length is the usual assumption. This is partly, because solely the general mathematical problem is considered, but often the local slip pattern is supposed to represent a superhydrophobic surface or a surface of specific morphology. 

For the mathematical modeling of specific physical surfaces, it is essential to distinguish the origin of the local slip. If the local slip is of intrinsic nature, it can be regarded as a material property. In this case, it may be isotropic. It can also be a constant and exhibit jumps at the edges of the material patches. If, in contrast, the local slip is an apparent slip due to a fluid enclosed in the surface corrugations, it is determined by the local hydrodynamics. In that case, it is a continuous function, which will usually be anisotropic, depending on the actual geometry. Of course, also combinations of these cases may occur.

So far, in the investigation of slip, usually only the fluid above the surface has been taken into account. When microstructured surfaces are addressed, where one fluid is in Cassie state, also the effects occurring "below the surface" should be considered.

\section{Conclusions}

This article is concerned with the modeling of flow past a periodic array of rectangular grooves that contain a second immiscible fluid. Such surfaces occur in numerous practical applications, for example when water flows along a superhydrophobic surface or for flow over porous media or membranes. Such flows scenarios are especially attractive technologically, as upon appropriate design they promise a significant reduction of the flow resistance of a surface and may therefore give rise to the development of energy-efficient surface structures.

Via a superposition technique, a model for transverse and longitudinal flow over such surfaces has been derived. The flow of the secondary fluid within the grooves is heuristically modeled and coupled to the flow over the surface by the local slip-length distribution. As a special case, the flow field over a single groove, which has been considered earlier \citep{schoenecker2013}, is exactly reproduced. So far, comparable expressions for the flow field over similar surfaces have only been derived for certain special cases like an infinite local slip or in terms of an infinite series for single-phase flow. In contrast, the present theory is a unified, closed-form description for two arbitrary Newtonian fluids. From the solution for the flow field, explicit analytical equations for the transverse and longitudinal effective slip length are derived. For the first time, they relate the effective slip length to the viscosities of the involved fluids, as well as to the geometry of the grooves. The equations for the longitudinal flow field and effective slip length are equally valid for single-phase flow, which might be of interest in aerodynamics. As comparisons with numerical calculations show, the derived equations are very accurate over the entire range of the governing parameters. In particular, these parameters are the viscosity ratio of the fluids, the aspect ratio of the grooves and the fluid-fluid interface fraction, the influence of which has been illustrated. Notably, no tunable fitting parameters are involved in the modeling, but the individual steps rely on physical argumentation. 

From the model equations, an expression is derived that relates the transverse effective slip length to that of a longitudinal flow over a similar surface. As a consequence, the mathematical effort when calculating effective slip lengths can be significantly reduced. In this context, it is essential that the local slip length of a grooved surface is in fact anisotropic. This point does not seem to have been acknowledged in the literature; however, it is important for the calculation of the effective slip length as well as for the design of realistic surfaces. Furthermore, the equations for the effective slip length allow an assessment of the common practice to model the air-water interface of a superhydrophobic surface as perfectly slipping. 

Overall, the equations presented could enable further investigations of slippage along structured surfaces. It is conceivable that a similar modeling strategy can also be applied to alternative geometries, like u-shaped or triangular grooves. Additionally, the presented method of superposition can be used to generalize results in other channel geometries, such as the results given by \citet{philip1972,philip1972b} for circular pipes or planar channels having walls with perfect slip and no-slip boundary conditions.

\section*{Acknowledgements}

We kindly acknowledge support by the Cluster of Excellence 259 'Center of Smart Interfaces' and by the Graduate School of Computational Engineering at TU Darmstadt, both funded by the German Research Foundation (DFG).

%%%%%%%%%%%%%%%%%%%%%%%%%%%%%%%%%%%%%%%%%%%%%%%%%%%%
%%%%%%%%%%%%%%%%%%%%%%%%%%%%%%%%%%%%%%%%%%%%%%%%%%
\appendix

%%%%%%%%%%%%%%%%%%%%%%%%%%%%%%%%%%%%%%%%%%%%%%%%%%%%%%%%%%%%%%%%%%%%%%%%%%%%%%%%%%%%
\section{Longitudinal flow over grooves with closed ends}

The introduced modeling procedure also allows considering different groove geometries, which leaves room for various future investigations. One especially relevant case is briefly illustrated here: longitudinal flow over grooves, whose ends are closed. In microfluidic applications, one will preferentially investigate longitudinal configurations of grooves, since they promise more slip than transverse configurations. However, it will not always be possible or desirable to allow for an in- and outflow of fluid in the grooves. This would in fact be necessary to assure a flow of the groove-filling medium in only one direction. If, in contrast, the fluid is contained in grooves with closed ends, there will always have to be a backflow due to mass conservation. Obviously, this backflow reduces the maximum slip to be expected compared to unidirectional flow. Hence, this scenario is of great practical interest, but has not been treated theoretically so far.

As a basic assumption, the grooves are considered to be much longer than wide, so that the effects at their ends may be neglected and a lubrication approximation can be applied. From this, it follows that in the limit of $A\rightarrow0$, the slope of the maximum local slip length $D_\mr {l,\,b}$ for longitudinal flow with backflow behaves like in the transverse case.

Therefore, the maximum local slip length for longitudinal flow over grooves with closed ends can be written analogously to the transverse case as
\begin{equation}
D_\mr{l,\,b}(A,a)=f(a)\, \hat D_{\mr{ l,\,b}}\,\erf\lk(\frac{g(a)\sqrt{\pi}}{8 f(a)\, D_{\mr{ l,\,b}}}\,A\rk).
\label{eq:dlbp}
\end{equation}
The prefactor can be described by
\begin{equation}
\hat D_{\mr{ l,\,b}}=\beta_{\mr{ l,\,N=1,\,max}}^*-\frac{1}{2}\lk(\beta_{\mr{ l,\,N=1,\,max}}^*-\beta_{\mr{ t,\,N=1,\,max}}^*\rk) \exp^{-f(a) \beta_{\mr{ l,\,N=1,\,max}}^* \, A}.
\label{eq:faktorendenzu}
\end{equation}
In the previous cases, the prefactor was the normalized maximum effective slip length at a viscosity ratio of unity, i.e. $\beta_{\mr{ t,\,N=1,\,max}}^*$ and $\beta_{\mr{ l,\,N=1,\,max}}^*$, respectively. Here, for $A\rightarrow 0$, the prefactor is just the arithmetic mean of these two values, while for $A\rightarrow\infty$ the grooves are so deep that they can accommodate any backflow. Hence, the prefactor becomes identical to that of longitudinal flow with open ends. The description of the local slip length \eqref{eq:dlbp} and \eqref{eq:faktorendenzu} can simply be employed as an input to the equations for the longitudinal flow and effective slip length.

\begin{figure}%
\centering
\includegraphics[width=0.5\textwidth]{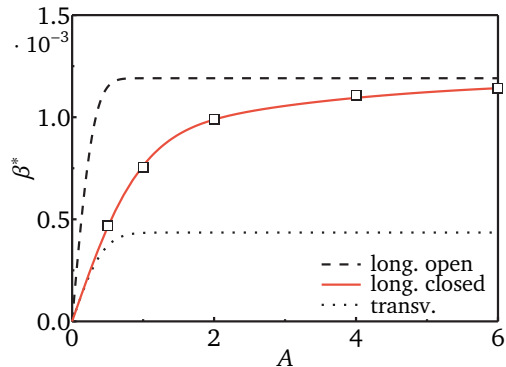}%
\caption{Dependence of the effective slip length on the aspect ratio. Comparison of longitudinal flow over grooves with open ends (eq. \eqref{eq:betaledc} with local slip \eqref{eq:dlp}), closed ends (eq. \eqref{eq:betaledc} with local slip \eqref{eq:dlbp}) and transverse flow (eq. \eqref{eq:betated} with local slip \eqref{eq:dtp}). Air is flowing over water-filled grooves ($N\approx 0.02$) with $a=0.5$. Squares indicate numerical results for longitudinal flow with closed grooves.}%
\label{fig:slipvgl_ltoffenzu}%
\end{figure}

Overall, the effective slip length for longitudinal flow over grooves with closed ends approaches its maximum at much larger values of $A$ than the same flow over open grooves. This can be seen in fig.~\ref{fig:slipvgl_ltoffenzu} where the effective slip length for these two scenarios is compared, using the example of air flowing over water-filled grooves ($N\approx 0.02$). 
For shallow grooves, the longitudinal slip over closed grooves and the transverse slip are nearly identical. All these points need to be taken into account when designing microstructured surfaces for slip applications. 

Fig. \ref{fig:slipvgl_ltoffenzu} also shows very good agreement of the analytical slip calculation with numerical results. For other viscosity ratios, the agreement between analytical and numerical calculations is equally good. The numerical computations were again performed with Comsol as a two-dimensional problem in the $x$-$y$~plane. The condition for backflow is a mass balance in the $z$~direction in the sense of lubrication theory, requiring the flow into and out of the $x$-$y$~plane to be equal. This condition was implemented as a global constraint $\int_\mr{groove} w_2\dd x \dd y =0$
, which includes the consideration of the pressure gradient inside the grooves.

%%%%%%%%%%%%%%%%%%%%%%%%%%%%%%%%%%%%%%%%%%%%%%%%%%
\bibliographystyle{jfm}

\bibliography{sliplength}

\end{document}